\documentclass[reprint,twocolumn,aps,prc,preprintnumbers,superscriptaddress,nofootinbib,bibnotes,floatfix]{revtex4-2}

\usepackage{graphicx}
\usepackage{dcolumn}
\usepackage{bm}
\usepackage{hyperref}
\usepackage[mathlines]{lineno}
\usepackage{float}
\usepackage{caption}
\usepackage{subcaption}

\usepackage{amsmath, fourier}
\usepackage{enumerate}

\begin{document}

\preprint{APS/123-QED}

\title{Nuclear Shell Structure in a {\bf Finite-Temperature} Relativistic Framework}

\author{Herlik Wibowo}
 \affiliation{Institute of Physics, Academia Sinica, Taipei, 11529, Taiwan}
\author{Elena Litvinova}%

\affiliation{Department of Physics, Western Michigan University, Kalamazoo, Michigan 49008, USA}
\affiliation{National Superconducting Cyclotron Laboratory, Michigan State University, East Lansing, Michigan 48824, USA}




\date{\today}

\begin{abstract}
The shell evolution of neutron-rich nuclei with temperature is studied in a beyond-mean-field framework rooted in the meson-nucleon Lagrangian.
The temperature-dependent Dyson equation with the dynamical kernel taking into account the particle-vibration coupling (PVC) is solved for the fermionic propagators in the basis of the thermal relativistic mean-field Dirac spinors. The calculations are performed for $^{68-78}$Ni in a broad range of temperatures $0 \leq T \leq 4$ MeV.  
The special focus is put on the fragmentation pattern of the single-particle states, which is further investigated within toy models in strongly truncated model spaces. Such models allow for quantifying the sensitivity of the fragmentation to the phonon frequencies, the PVC strength and to the mean-field level density. The model studies provide insights into the temperature evolution of the PVC mechanism in real nuclear systems under the conditions which may occur in astrophysical environments. 
\end{abstract}

\maketitle



\section{\label{sec:level1}Introduction}

Understanding the behavior of atomic nuclei and nuclear matter at finite temperature is extremely important for advancements at the frontiers of the nuclear science. 
The modification of in-medium nucleonic correlations with temperature changes considerably the nuclear structure, leading to the transition to the non-superfluid phase, weakening of collective effects, the appearance of shape fluctuations, and the formation of new structures in the excitation spectra due to the thermal unblocking \cite{AlhassidBushLevit1988,AlhassidBush1990c,Kusnezov1998,Litvinova2018d,Wibowo2019a,LitvinovaRobinWibowo2020}. The microscopic interpretation of these phenomena is crucial for accurate predictions of nuclear processes in astrophysical environments, such as the neutron star mergers and supernovae \cite{MumpowerSurmanMcLaughlinEtAl2016,Lattimer2019,Raduta2017,Nagakura2019,Pascal2020,Langanke2021}. As it is found in the recent studies of Refs. \cite{LitvinovaRobinWibowo2020,Litvinova2021}, the emergent effects of the many-body correlations, of both collective and non-collective origin, play a decisive role in the key nuclear reaction rates employed in the r-process nucleosynthesis and core-collapse supernovae simulations. The precise knowledge about the evolution of  nuclear emergent phenomena with temperature is, therefore, mandatory for a high-quality nuclear physics input utilized in astrophysical modeling.

The equation of motion (EOM) method builds a systematic framework for the description of many-body correlations, in particular, in strongly coupled fermionic systems
 \cite{Tiago2008,Martinez2010,Sangalli2011,SchuckTohyama2016,Olevano2018,Schuck2021}.  Within this framework, the EOM's are generated for various time-dependent quantities, such as the correlation functions of field operators. One of the simplest correlation functions is the fermionic propagator through the correlated medium, which is directly related to the energies of  quasiparticles and their occupancies of the basis orbitals \cite{Dickhoff2005,Dickhoff2004}. In principle, the single-particle propagators of the states below the Fermi energy define completely the total ground state energy of the system, if the underlying Hamiltonian is confined by the two-body interaction, 
via the Midgal-Galitski-Koltun sum rule \cite{GalitskiMigdal1958,Koltun1974}. This fact is, in turn, in compliance with the density functional theory, where the total ground state energy is defined by the one-body density, which is the static limit of the single-particle propagator. Thus, the correlated one-body propagator plays a fundamental role in the description of quantum many-body systems.

However, the exact EOM for the one-body propagator does not have a closed form. Instead, it contains higher-rank propagators in the dynamical part of the interaction kernel, namely, the two-body propagator in the non-symmetric form and the three-body propagator in the symmetric form of this kernel \cite{LitvinovaSchuck2019,Litvinova2021a}. This requires external EOMs for the latter propagators, which, in turn, are coupled to even higher-rank ones through the more complex dynamical kernels, leading to a hierarchy of coupled EOMs. In nuclear physics applications, however, quantitatively most important coupling is the one between the one-body and two-body propagators, while the two-body EOM can be formulated in various approximations, which allow one to truncate the hierarchy of EOMs on a certain level with a reasonable accuracy. 

The simplest descriptions of quantum many-body systems take the dynamical kernels of the EOMs into account in static approximations. Such descriptions are confined by one-body densities and propagators and include the Hartree-Fock approach \cite{Hartree1928,Fock1930,HF1935}, the random phase approximation (RPA) \cite{Bohm1951}, 
the Gor'kov theory of superfluidity \cite{Gorkov1958} and the Bardeen-Cooper-Schrieffer (BCS) model \cite{Bardeen1957}, to name a few. Needless to say, such approximations neglect the explicit coupling between the one-body and two-body EOM's and, thus, are insufficient for an accurate description of nuclear phenomena, which is required for modern applications and for a deep understanding of emergent effects.
A  better accuracy can be achieved by cluster expansions of the dynamical kernels of both the one-body and two-body fermionic EOM's in terms of the two-time two-body correlation functions, as it is discussed, e.g.,  in the context of the condensed matter and quantum chemistry applications \cite{Tiago2008,Martinez2010,Sangalli2011,Olevano2018}. For the nuclear physics calculations, possible truncation schemes of the dynamical kernels on the two-body level were discussed, in particular, in Ref. \cite{LitvinovaSchuck2019}.  The great advantage of the EOM method is that  both the static and dynamical kernels are derived consistently from the same underlying bare interaction. In nuclear physics, however, the implementations of EOM-based methods still mainly employ effective interactions. 

Early approaches with dynamical kernels postulated phenomenological Hamiltonians implying the existence of fermionic quasiparticles and phonons of a bosonic origin  \cite{BohrMottelson1969,BohrMottelson1975,BortignonBrogliaBesEtAl1977,Broglia1981,BertschBortignonBroglia1983,Soloviev1992,Ponomarev2001,SavranBabilonBergEtAl2006,Andreozzi2008}.  In such approaches the effective residual interaction between fermions is supplemented by the phonon-exchange interaction, or (quasi)particle-vibration coupling ((q)PVC). Relatively simple calculation schemes are possible with the use of effective phenomenological interactions, however, more accurate and sophisticated versions of PVC were successfully implemented over the years \cite{IdiniBarrancoVigezzi2012,Litvinova2016,RobinLitvinova2016,RobinLitvinova2018,Niu2018,Robin2019,Tselyaev2018,Lyutorovich:2018cph,LitvinovaSchuck2019,Shen2020}. The models operating with mostly the phonon degrees of freedom \cite{Soloviev1992,Ponomarev2001,SavranBabilonBergEtAl2006,Andreozzi2008}, include  correlations of high complexity, and a few attempts of using  bare nucleon-nucleon interactions were reported \cite{DeGregorio2017,DeGregorio2017a,Knapp:2014xja,Knapp:2015wpt}.  

In this work we examine the EOM for the one-body propagator, which is, in its most general form, the Dyson equation. We consider the finite-temperature case and employ the approach developed in Ref. \cite{Wibowo2020} as an extension of the zero-temperature PVC model  \cite{LitvinovaRing2006,Litvinova2012}. The latter are built upon the relativistic quantum hadrodynamics \cite{SerotWalecka1986a,SerotWalecka1979,Ring1996,VretenarAfanasjevLalazissisEtAl2005} adopting the  
PVC mechanism \cite{RingSchuck1980,DukelskyRoepkeSchuck1998,LitvinovaSchuck2019,Schuck2019} for the induced interaction. As it follows from Refs. \cite{DukelskyRoepkeSchuck1998,LitvinovaSchuck2019}, PVC is the leading contribution to the one-fermion dynamical kernel, or the self-energy,  
in finite nuclei. Formally similar to the phenomenological PVC proposed quite early by A. Bohr and B. Mottelson \cite{BohrMottelson1969,BohrMottelson1975}, it is now understood in terms of the EOM derived from ab-initio nucleon-nucleon potentials \cite{LitvinovaSchuck2019}. Although in this work we still keep the phenomenological effective interaction adjusted in the framework of the covariant density functional theory (CDFT) \cite{VretenarAfanasjevLalazissisEtAl2005} for the static part of the EOM kernel and the PVC, that allows one to obtain the PVC vertices and phonon propagators within the relativistic random phase approximation (RRPA) with a good accuracy. This feature remains intact also at finite temperature, which is another important ingredient of our study. Furthermore, we focus on a systematic application of the approach to the temperature evolution of the PVC mechanism in the neutron-rich even-even nuclei $^{68-78}$Ni, which are of interest for various astrophysical applications.

\section{Dyson Equation for the Fermionic Propagator at Finite Temperature}

We define the hot nucleus as a system of Dirac nucleons moving in a self-consistent 
field generated by an effective meson-nucleon interaction at finite temperature. The electromagnetic interaction between protons is mediated by the photon. 
One of the most convenient ways to quantify the single-particle motion in a fermionic many-body system is  to evaluate the one-fermion propagator (also called Green function), which describes the motion of a fermion through the correlated medium formed by $N$ identical interacting fermions. The advantage of such a description is the simple relationship of this propagators to the excitation spectra and ground-state properties of the systems with $(N+1)$ and $(N-1)$ fermions. In this work we are interested in nuclear systems in thermal equilibrium with the surroundings, that can be associated with a certain temperature. The temperature, or Matsubara, propagator of a fermion is defined as \cite{Matsubara1955, Abrikosov1975, Zagoskin2014}
\begin{equation}
\label{Matsubara Green function}{\cal G}(1,1')\equiv{\cal G}_{k_{1}k_{1'}}(\tau_{1}-\tau_{1'})=-\langle T_{\tau}\psi(1)\overline{\psi}(1')\rangle,
\end{equation}
where the angular brackets stand for the thermal average \cite{Abrikosov1975, Zagoskin2014} and the chronological ordering operator $T_{\tau}$ acts on the fermionic field operators in the Wick-rotated picture:
\begin{eqnarray}
\psi(1)&\equiv &\psi_{k_{1}}(\tau_{1})=e^{{\cal H}\tau_{1}}\psi_{k_{1}}e^{-{\cal H}\tau_{1}},\nonumber\\
\overline{\psi}(1)&\equiv&\psi^{\dag}_{k_{1}}(\tau_{1})=e^{{\cal H}\tau_{1}}\psi^{\dag}_{k_{1}}e^{-{\cal H}\tau_{1}}.
\label{FOs}
\end{eqnarray}
In Eq. (\ref{FOs}), the evolution is determined by the operator ${\cal H}=H-\mu N$, where $H$ is the many-body Hamiltonian, $\mu$ is the chemical potential, and $N$ denotes the particle number. The subscript $k_{1}$ defines the full set of the single-particle quantum numbers in a given representation, while the imaginary time variable $\tau$ is related to the real time $t$ as $\tau=it$. The fermionic fields $\psi_{k_{1}}$ and $\psi^{\dagger}_{k_{1}}$ satisfy the usual anticommutation relations. 

For the many-body Hamiltonian $H$ containing only the free-motion and the mean-field contributions,  i.e., confined by the one-body part, the single-fermion Matsubara Green function reads
\begin{eqnarray}
\widetilde{{\cal G}}(2,1)&=&\sum_{\sigma=\pm 1}\widetilde{{\cal G}}^{\sigma}(2,1),\nonumber\\
\label{Thermal G}\widetilde{{\cal G}}^{\sigma}(2,1)&=&-\sigma\delta_{k_{2}k_{1}}n(-\sigma(\varepsilon_{k_{1}}-\mu),T)e^{-(\varepsilon_{k_{1}}-\mu)\tau_{21}}\theta(\sigma\tau_{21}),\nonumber\\
\label{MFGF}
\end{eqnarray}
where $\tau_{21}=\tau_{2}-\tau_{1}$ and $\varepsilon_{k_{1}}$ are the eigenvalues of the single-particle Hamiltonian diagonal in the $\{k_{i}\}$ basis. Accordingly,  $n(\varepsilon,T)$ is the Fermi-Dirac distribution, 
\begin{equation}
\label{Thermal occupation numbers}n(\varepsilon,T)=\frac{1}{\exp(\varepsilon/T)+1},
\end{equation} 
at the temperature $T$ and characterizes the mean-field occupancies of the orbits with the given single-particle energies. The Fourier transform of the propagator (\ref{MFGF}) to the energy domain, 
\begin{equation}
\widetilde{{\cal G}}_{k_{2}k_{1}}(\varepsilon_{\ell})=\int_{0}^{1/T}d\tau e^{i\varepsilon_{\ell}\tau}\widetilde{{\cal G}}_{k_{2}k_{1}}(\tau),
\end{equation}
leads to its spectral representation
\begin{equation}
\label{Spectral Termal G}\widetilde{\cal G}_{k_{2}k_{1}}(\varepsilon_{\ell})=\delta_{k_{2}k_{1}}\widetilde{\cal G}_{k_{1}}(\varepsilon_{\ell}),\;\;\;\;\;\widetilde{\cal G}_{k_{1}}(\varepsilon_{\ell})=\frac{1}{i\varepsilon_{\ell}-\varepsilon_{k_{1}}+\mu},
\end{equation}
defined at the discrete Matsubara frequencies $\varepsilon_{\ell}$,
\begin{equation}
\varepsilon_{\ell}=(2\ell+1)\pi T,
\end{equation}
with the integer $\ell$. In Eqs. \eqref{Thermal G}-\eqref{Spectral Termal G} we indicate the mean-field character of the respective Green function by the " $\widetilde{}$ " sign. 

The presence of two-body and higher-rank terms in the many-body Hamiltonian induce correlations beyond mean field originated  from the residual interaction. The correlated propagator can be found as a solution of  the Dyson equation
\begin{equation}
\label{Dyson Equation in terms of thermal G}{\cal G}_{k_{1}k_{2}}(\varepsilon_{\ell})=\widetilde{{\cal G}}_{k_{1}k_{2}}(\varepsilon_{\ell})+\sum_{k_{3}k_{4}}\widetilde{{\cal G}}_{k_{1}k_{3}}(\varepsilon_{\ell})\Sigma^{e}_{k_{3}k_{4}}(\varepsilon_{\ell}){\cal G}_{k_{4}k_{2}}(\varepsilon_{\ell}),
\end{equation}
where the energy-dependent mass operator, or self-energy, $\Sigma^{e}$ describes the coupling between single fermions and in-medium emergent degrees of freedom. In this work, we employ the PVC model for $\Sigma^{e}$, which approximates the exact energy-dependent self-energy $\Sigma^{e}$ by a cluster expansion truncated at the two-body level \cite{LitvinovaSchuck2019}. This self-energy, in the leading approximation, reads
\begin{equation}
\label{Construction of Sigma e}\Sigma^{e}_{k_{1}k_{2}}(\varepsilon_{\ell})=-T\sum_{k_{3},m}\sum_{\ell'}\sum_{\sigma=\pm 1}\widetilde{\cal G}_{k_{3}}(\varepsilon_{\ell'})\frac{\sigma g_{k_{1}k_{3}}^{m(\sigma)}g_{k_{2}k_{3}}^{m(\sigma)\ast}}{i\varepsilon_{\ell}-i\varepsilon_{\ell'}-\sigma\omega_{m}},
\end{equation}
where $g^{m}$ are the phonon vertices and $\omega_{m}$ are their frequencies. The vertices corresponding to the specific frequencies can be extracted from the EOM for the two-fermion propagators as follows:
\begin{eqnarray}
\label{Phonon vertices}g^{m}_{k_{1}k_{2}}=\sum_{k_{3}k_{4}}\widetilde{{\cal U}}_{k_{1}k_{4},k_{2}k_{3}}\rho^{m}_{k_{3}k_{4}},\\
g^{m(\sigma)}_{k_{1}k_{2}}=\delta_{\sigma,+1}g^{m}_{k_{1}k_{2}}+\delta_{\sigma,-1}g^{m}_{k_{2}k_{1}},
\end{eqnarray}
where $\rho^{m}_{k_{3}k_{4}}$ are the matrix elements of the transition density for the $m$-th mode of excitation of the $N$-particle system and $\widetilde{{\cal U}}_{k_{1}k_{4},k_{2}k_{3}}$ are the matrix elements of the nucleon-nucleon interaction. As shown in Ref. \cite{Litvinova2019g}, the relationship \eqref{Phonon vertices} is model independent and includes the exact transition densities, while the interaction $\widetilde{{\cal U}}$ is the bare interaction between nucleons in the vacuum. In practice, employing effective interactions and the random phase approximation based on these interactions for the computation of the phonon characteristics provide quite a realistic description of the dynamical self-energy. In this work, we use the effective interaction of the covariant energy density functional (CEDF) \cite{Ring1996, VretenarAfanasjevLalazissisEtAl2005} with the NL3 parametrization \cite{Lalazissis1997a} and the relativistic random phase approximation \cite{Ring2001a}  adopted to finite temperature in our previous developments for calculations of the phonon modes \cite{Litvinova2018d, Wibowo2019a, Litvinova2019g}.

The summation over $\ell'$ in Eq. \eqref{Construction of Sigma e} is transformed into a contour integral by the standard technique \cite{Zagoskin2014}. The final expression for the mass operator $\Sigma^{e}$, after the analytical continuation to complex energies, takes the form:
\begin{eqnarray}
\label{Sigma e}\Sigma^{e}_{k_{1}k_{2}}(\varepsilon)&=&\sum_{k_{3},m}\left\{g^{m}_{k_{1}k_{3}}g^{m\ast}_{k_{2}k_{3}}\frac{N(\omega_{m},T)+1-n(\varepsilon_{k_{3}}-\mu,T)}{\varepsilon-\varepsilon_{k_{3}}+\mu-\omega_{m}+i\delta}\right.\nonumber\\
&+&\left. g^{m\ast}_{k_{3}k_{1}}g^{m}_{k_{3}k_{2}}\frac{n(\varepsilon_{k_{3}}-\mu,T)+N(\omega_{m},T)}{\varepsilon-\varepsilon_{k_{3}}+\mu+\omega_{m}-i\delta}\right\},
\end{eqnarray}
where $\delta\rightarrow+0$, and 
\begin{equation}
N(\omega_{m},T)=\frac{1}{\exp(\omega_{m}/T)-1}
\end{equation}
are the occupation numbers of phonons with the frequencies $\omega_{m}$, which arise from the summation over $\ell'$ in Eq. (\ref{Construction of Sigma e}).

As in Ref. \cite{Wibowo2020}, in this work the self-energy (\ref{Sigma e}) is treated in the diagonal approximation: $\Sigma^{e}_{k_{1}k_{2}}(\varepsilon) = \delta_{k_{1}k_{2}}\Sigma^{e}_{k_1}(\varepsilon)$.  Furthermore, Eq. \eqref{Dyson Equation in terms of thermal G} is equivalent to the nonlinear equation:
\begin{equation}
\left[\varepsilon-\varepsilon_{k}+\mu-\Sigma^{e}_{k}(\varepsilon)\right]{\cal G}_{k}(\varepsilon)=1,
\end{equation}
for each single-particle state $k$. The poles of the propagator ${\cal G}_{k}(\varepsilon)$ correspond to the zeros of the function
\begin{equation}
f(\varepsilon)=\varepsilon-\varepsilon_{k}+\mu-\Sigma^{e}_{k}(\varepsilon).
\end{equation}
For each single-particle mean-field state $k$, there exist multiple solutions $\varepsilon_{k}^{(\lambda)}$ numbered by the additional index $\lambda$. In other words, the pole character of the energy-dependent self-energy $\Sigma^{e}_{k}(\varepsilon)$ causes fragmentation of the single-particle states $k$ due to the PVC mechanism. The solutions $\varepsilon_{k}^{(\lambda)}$ can be determined by finding the zeros of $f(\varepsilon)$ or, alternatively, by the diagonalization of the arrowhead matrix \cite{Ring1973, Litvinova2006a}:
\begin{equation}
\left( \begin{array}{cccc}
\varepsilon_{k}-\mu & \xi^{m_{1}(\sigma)}_{n_{1}k} & \xi^{m_{2}(\sigma)}_{n_{2}k} & \cdots \\ 
\xi^{m_{1}(\sigma)\ast}_{n_{1}k} & \varepsilon_{n_{1}}-\mu-\sigma\omega_{m_{1}} & 0 & \cdots \\ 
\xi^{m_{2}(\sigma)\ast}_{n_{2}k} & 0 & \varepsilon_{n_{2}}-\mu-\sigma\omega_{m_{2}} & \cdots \\ 
\vdots & \vdots & \vdots & \ddots
\end{array}  \right),
\end{equation}
where 
\begin{equation}
\xi^{m(\sigma)}_{nk}=g^{m(\sigma)}_{nk}\sqrt{N(\omega_{m},T)+n(\sigma(\varepsilon_{n}-\mu),T)}, \ \ \ \ \ \sigma=\pm 1.
\end{equation}
In the vicinity of the pole $\varepsilon_{k}^{(\lambda)}$, where the function $f(\varepsilon)$ can be approximated by
\begin{equation}
f(\varepsilon)\approx \left(\varepsilon-\varepsilon_{k}^{(\lambda)}\right)\left[1-\frac{d}{d\varepsilon}\Sigma^{e}_{k}(\varepsilon)\right]_{\varepsilon=\varepsilon^{(\lambda)}_{k}},
\end{equation}
the correlated Matsubara Green function ${\cal G}_{k}(\varepsilon)$ reads:
\begin{equation}
{\cal G}_{k}^{(\lambda)}(\varepsilon)\approx\frac{S^{(\lambda)}_{k}}{\varepsilon-\varepsilon_{k}^{(\lambda)}},
\end{equation}
with the spectroscopic factor $S^{(\lambda)}_{k}$, such as
\begin{equation}
\label{Spectroscopic factor}S_{k}^{(\lambda)}=\left[1-\frac{d}{d\varepsilon}\Sigma^{e}_{k}(\varepsilon)\right]^{-1}_{\varepsilon=\varepsilon^{(\lambda)}_{k}}.
\end{equation}
The spectroscopic factor $S^{(\lambda)}_{k}$ provides a measure for the occupancy of the state $\lambda$ with the single-particle quantum number $k$. The spectroscopic factors $S^{(\lambda)}_{k}$ and the energies of the correlated states $\varepsilon_{k}^{(\lambda)}$ satisfy the well-known sum rules \cite{Birbrair2000}:
\begin{equation}
\sum_{\lambda}S^{(\lambda)}_{k}=1 \ \ \ \ \ \ \ \ \ \ \ \ \sum_{\lambda}\varepsilon_{k}^{(\lambda)}S^{(\lambda)}_{k} = \varepsilon_{k},
\label{Sum Rule}
\end{equation}
which remain valid at finite temperature.

\section{Numerical Scheme}
The numerical implementation is performed in three steps. (i) The closed set of the relativistic mean-field (RMF) equations with the NL3 parametrization \cite{Lalazissis1997a} and the thermal fermionic occupation numbers \eqref{Thermal occupation numbers} is solved in a self-consistent cycle. This procedure outputs the single-particle spectrum as a set of temperature-dependent single-particle Dirac spinors and the corresponding energies, which form the basis $\{k_i\}$ employed in further calculations. (ii) The finite-temperature RRPA (FT-RRPA) equations are solved to obtain the phonon vertices $g^{m}$ and their frequencies $\omega_{m}$. The  FT-RRPA phonon spectrum with the RMF single (quasi)particles, build the $pp\otimes\text{phonon}$ and $ph\otimes\text{phonon}$ configurations for the PVC self-energy $\Sigma^{e}(\varepsilon)$. (iii) Eq. \eqref{Dyson Equation in terms of thermal G} is solved in the configuration space, truncated as described in \cite{Wibowo2020}. The PVC self-energy $\Sigma^{e}(\varepsilon)$ is treated in the diagonal approximation, i.e., $\Sigma^{e}_{k_{1}k_{2}}(\varepsilon)=\delta_{k_{1}k_{2}}\Sigma^{e}_{k_{1}}(\varepsilon)$. Pairing correlations at $T = 0$ are taken into account in all the three steps: in the Bardeen-Cooper-Schrieffer (BCS) approximation \cite{Bardeen1957} for the mean-field calculations, in the relativistic quasiparticle RPA (RQRPA) \cite{PaarRingNiksicEtAl2003} for the calculations of the phonon spectra, and in the approach of Ref. \cite{Litvinova2012} for the solution of the Dyson equation. Pairing correlations were neglected in calculations at temperature $T \geq 1$ MeV as the critical temperature of the superfluid phase transition is around 1 MeV for the considered nuclei.

The particle-hole ($ph$) configurations with the energies $\varepsilon_{ph}\leq 100$ MeV and the antiparticle-hole ($\alpha h$) ones with $\varepsilon_{\alpha h}\geq-1800$ MeV, with respect to the positive-energy continuum,  were included in the particle-hole basis for the FT-RRPA and RQRPA calculations of the phonon spectra. The excitation spectra converge reasonably well with this truncation, as it was verified by direct calculations within the complete RMF basis. The resulting vibrations with the spin-parities $J^{\pi}=2^{+},\;3^{-},\;4^{+},\;5^{-},\;6^{+}$ below the energy cutoff of 20 MeV formed the phonon model space. This cutoff is also justified by our previous calculations \cite{Wibowo2020}. A further truncation of the phonon space was made according to the values of the reduced transition probabilities of the corresponding electromagnetic transitions $B(EL)$. Namely, the phonon modes with the $B(EL)$ values equal or more than 5\% of the maximal one for each $J^{\pi}$ were retained in the self-energy $\Sigma^{e}(\varepsilon)$. The same truncation criteria were applied for all temperature regimes. As in our previous calculations \cite{Litvinova2018d, Wibowo2019a, Litvinova2019g, Wibowo2020}, at high temperatures many additional phonon modes appear in the excitation spectra as a consequence of the thermal unblocking, which leads to a significant expansion of the phonon model space with the temperature growth. The single-particle intermediate states $k_{3}$ were included in the summation of Eq. \eqref{Sigma e}  under the condition $|\varepsilon_{k_{3}}-\varepsilon_{k_{1}}|\leq 50$ MeV. The latter implies another truncation of the model space, which is mild enough that the results converge. In contrast to the number of the phonon modes, the number of the intermediate fermionic states changes only little with temperature. 

\section{Results and Discussion}

The numerical implementation was performed for the chain of neutron-rich even-even Ni isotopes with the atomic numbers $A = 68-78$, which lie on the r-process path and belong to the high-sensitivity region of the nuclear landscape with respect to the electron capture \cite{MumpowerSurmanMcLaughlinEtAl2016,Lattimer2019,Raduta2017,Nagakura2019,Pascal2020,Langanke2021}. We investigated the thermal evolution of the single-particle states located within the 20 MeV energy window around the respective Fermi energies of the neutron and proton subsystems.  Among the selected isotopes,  $^{78}\text{Ni}$ is a closed-shell nucleus as well as the proton subsystems of the other five Ni isotopes. The neutron subsystems of $^{68,70,72,74,76}$Ni are open-shell. As discussed in detail in Refs. \cite{Wibowo2020,Litvinova2021b}, the superfluidity in the Bogoliubov's or BCS sense vanishes at the critical temperature $T_{c}\approx 0.6\Delta_{p}(T=0)$, where $\Delta_{p}(T=0)$ is the pairing gap $\Delta_{p}$ at zero temperature. In the realistic self-consistent finite-temperature calculations the coefficient between $T_{c}$ and $\Delta_{p}(T=0)$ can be slightly different, for instance, in the relativistic RMF+BCS calculations it can get close to 0.7 \cite{Wibowo2020,Litvinova2021b}. In the approaches beyond BCS the critical temperature further increases as shown, in particular, in Ref. \cite{Litvinova2021b}. In this work, however, we stay within the RMF+BCS description, where the critical temperature is around or below 1 MeV for the selected nuclei. Their zero-temperature neutron pairing gaps $\Delta_{p}(T=0)$ were determined using the three-point formula \cite{Broglia2013} and the data on the nuclear binding energies from Ref. \cite{Audi2003}. The empirical values of the pairing gap $\Delta_{p}(T=0)$ and the corresponding empirical critical temperature $T_{c}$ for the even-even $^{68-76}\text{Ni}$ nuclei are shown in Table \ref{Neutron pairing gaps and critical T}. Since our results below are presented on 1 MeV temperature grid, the pairing correlations are taken into account only at $T=0$.  

\begin{table}[h]
\caption{The zero-temperature pairing gaps $\Delta_{p}(T=0)$ and critical temperatures $T_{c}$ for even-even $^{68-76}\text{Ni}$ nuclei.}
\begin{ruledtabular}
\begin{tabular}{cccccc}
& $^{68}\text{Ni}$ & $^{70}\text{Ni}$ & $^{72}\text{Ni}$ & $^{74}\text{Ni}$ & $^{76}\text{Ni}$ \\ 
\hline 
$\Delta_{p}(T=0)$ (MeV) & 1.57 & 1.55 & 1.41 & 1.49 & 1.30 \\ 
\hline
$T_{c}$ (MeV) & 0.94 & 0.93 & 0.85 & 0.90 & 0.78 \\ 
\end{tabular} 
\end{ruledtabular}
\label{Neutron pairing gaps and critical T}
\end{table}


\begin{figure*}[h!]
\begin{subfigure}{0.4\textwidth}
\centering
\includegraphics[scale=0.35]{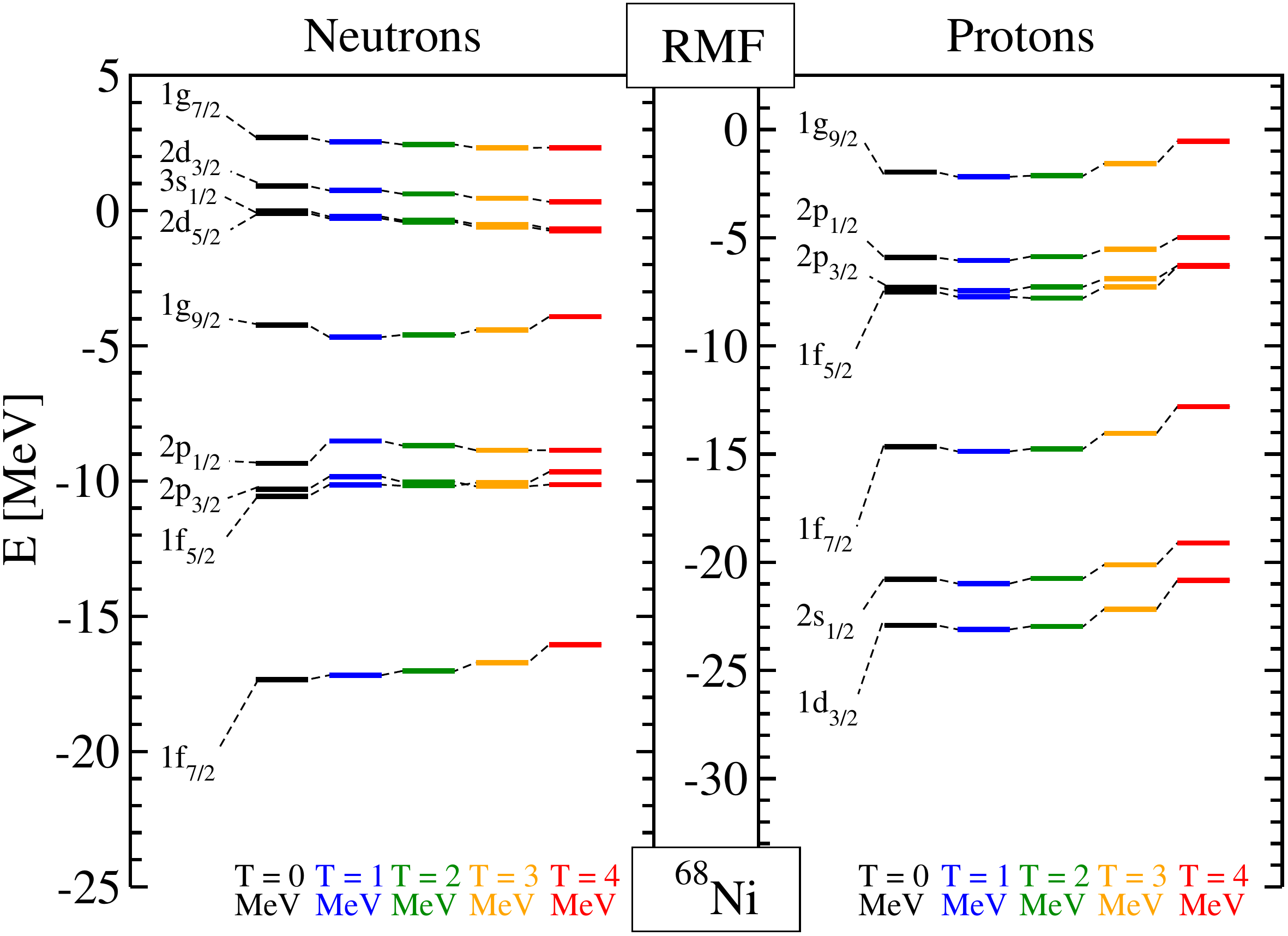} 
\caption{}
\label{Ni68_RMF} 
\end{subfigure}
\hfill
\begin{subfigure}{0.5\textwidth}
\centering
\includegraphics[scale=0.35]{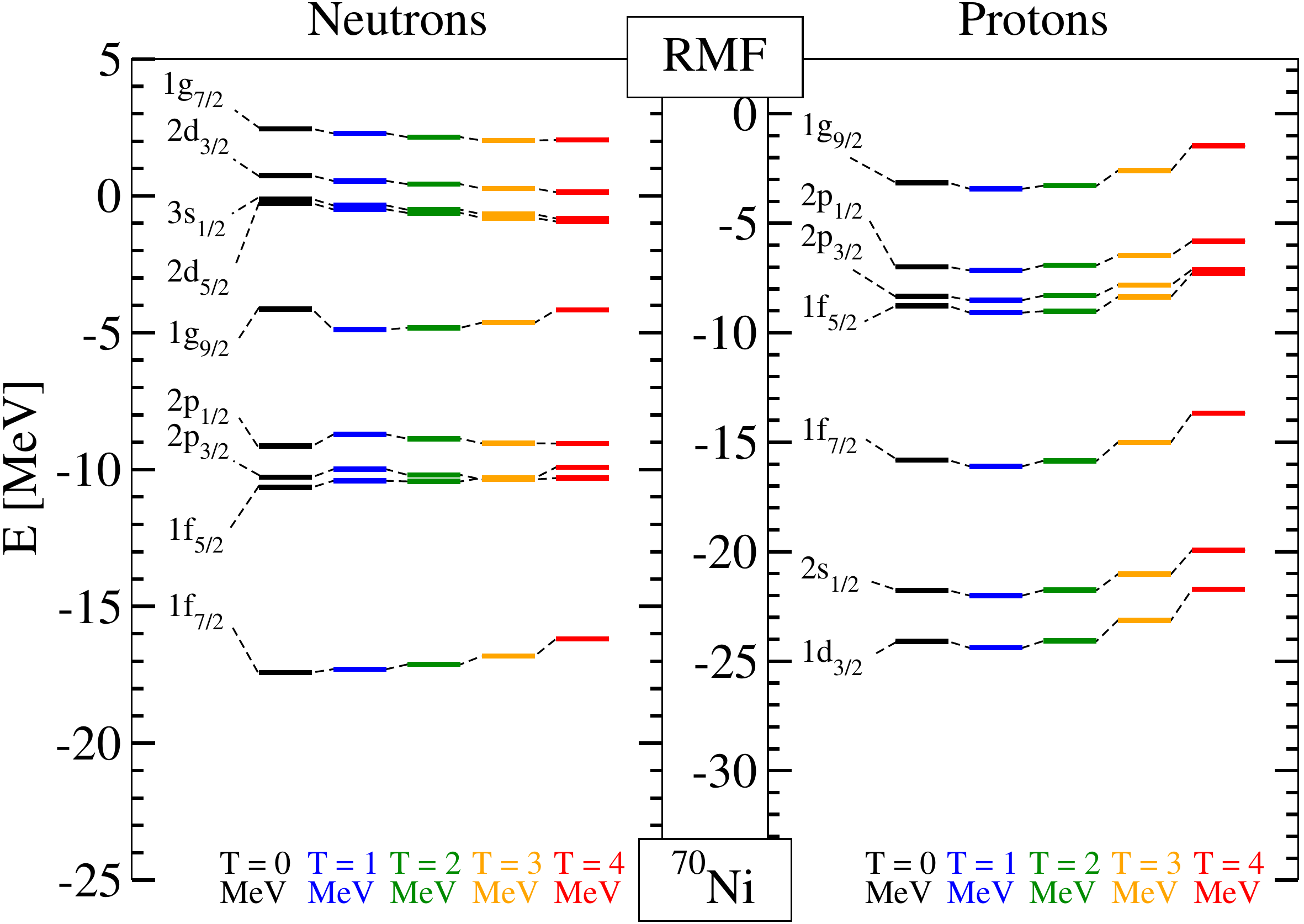}
\caption{}
\label{Ni70_RMF}
\end{subfigure}
\hfill
\begin{subfigure}{0.4\textwidth}
\centering
\includegraphics[scale=0.35]{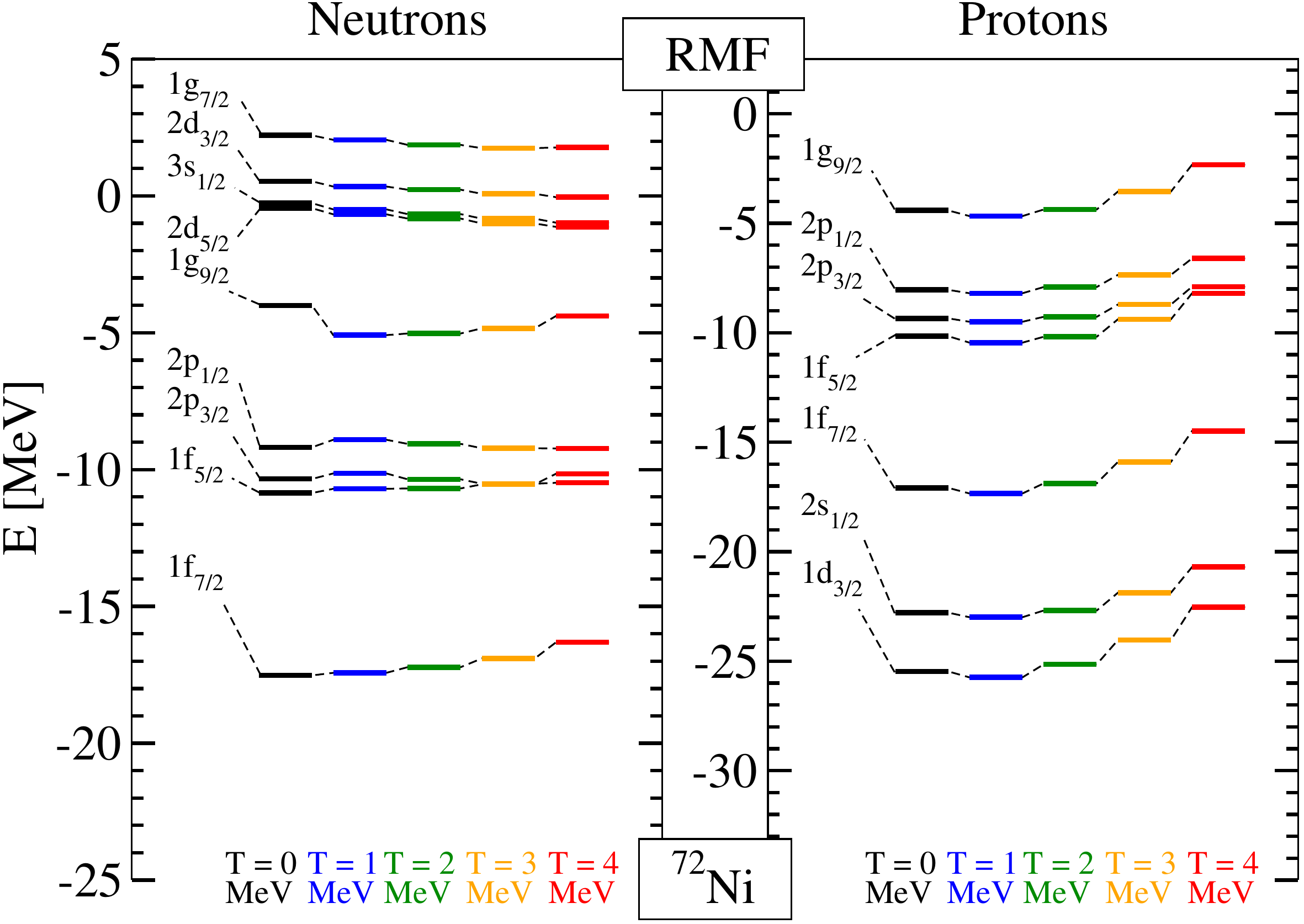}
\caption{}
\label{Ni72_RMF} 
\end{subfigure}
\hfill
\begin{subfigure}{0.5\textwidth}
\centering
\includegraphics[scale=0.35]{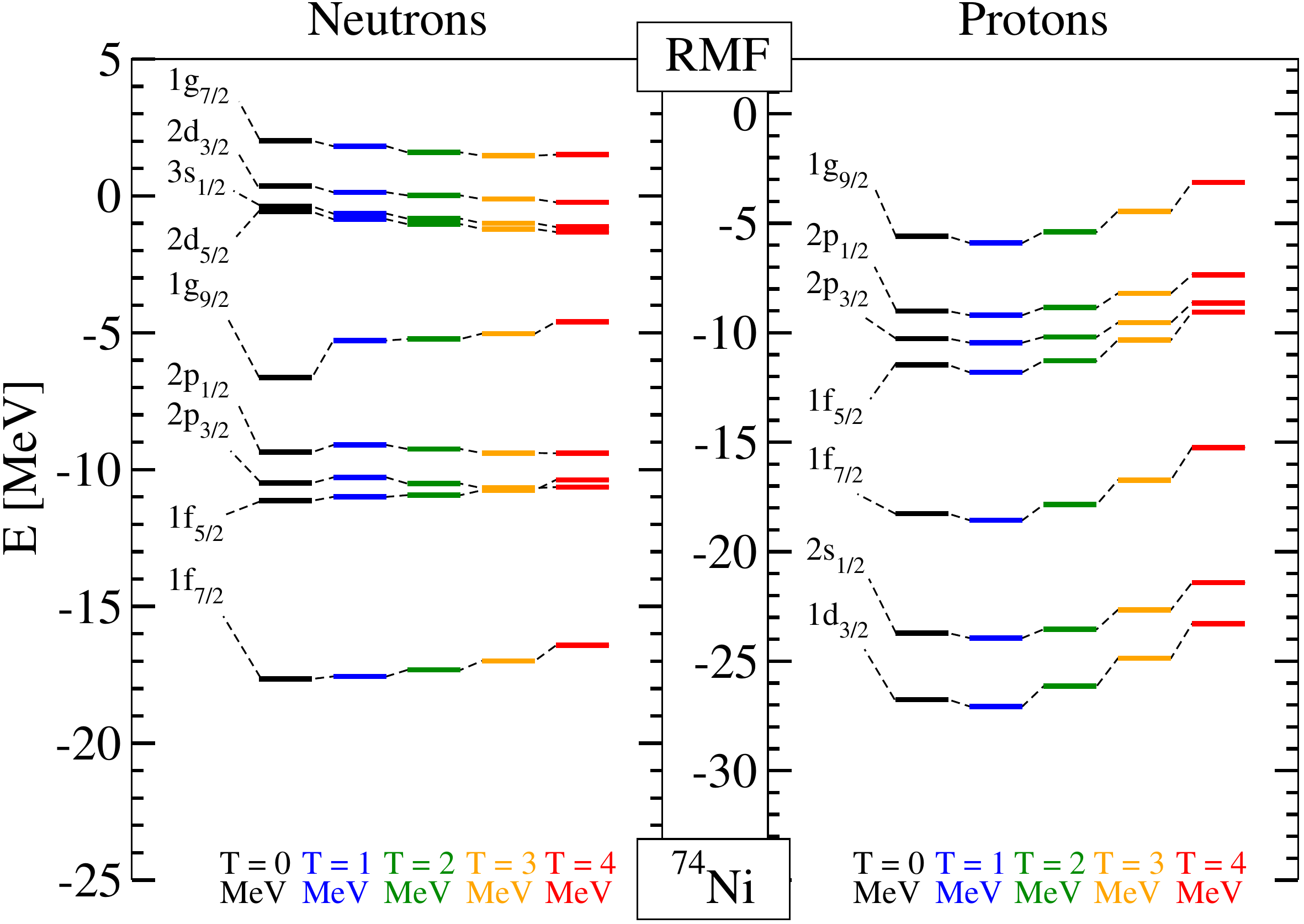}
\caption{} 
\label{Ni74_RMF} 
\end{subfigure}
\hfill
\begin{subfigure}{0.4\textwidth}
\centering
\includegraphics[scale=0.35]{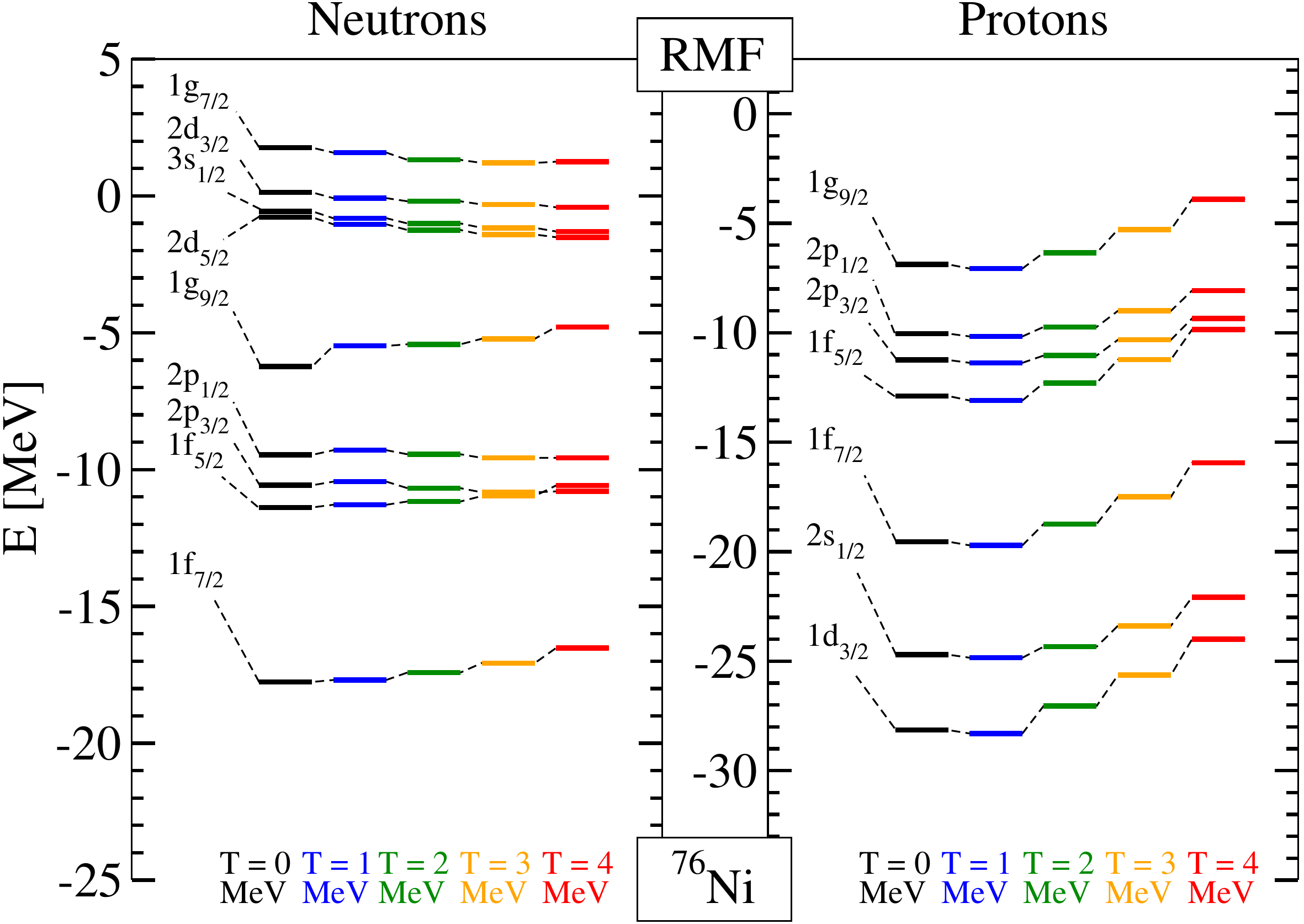}
\caption{} 
\label{Ni76_RMF} 
\end{subfigure}
\hfill
\begin{subfigure}{0.5\textwidth}
\centering
\includegraphics[scale=0.35]{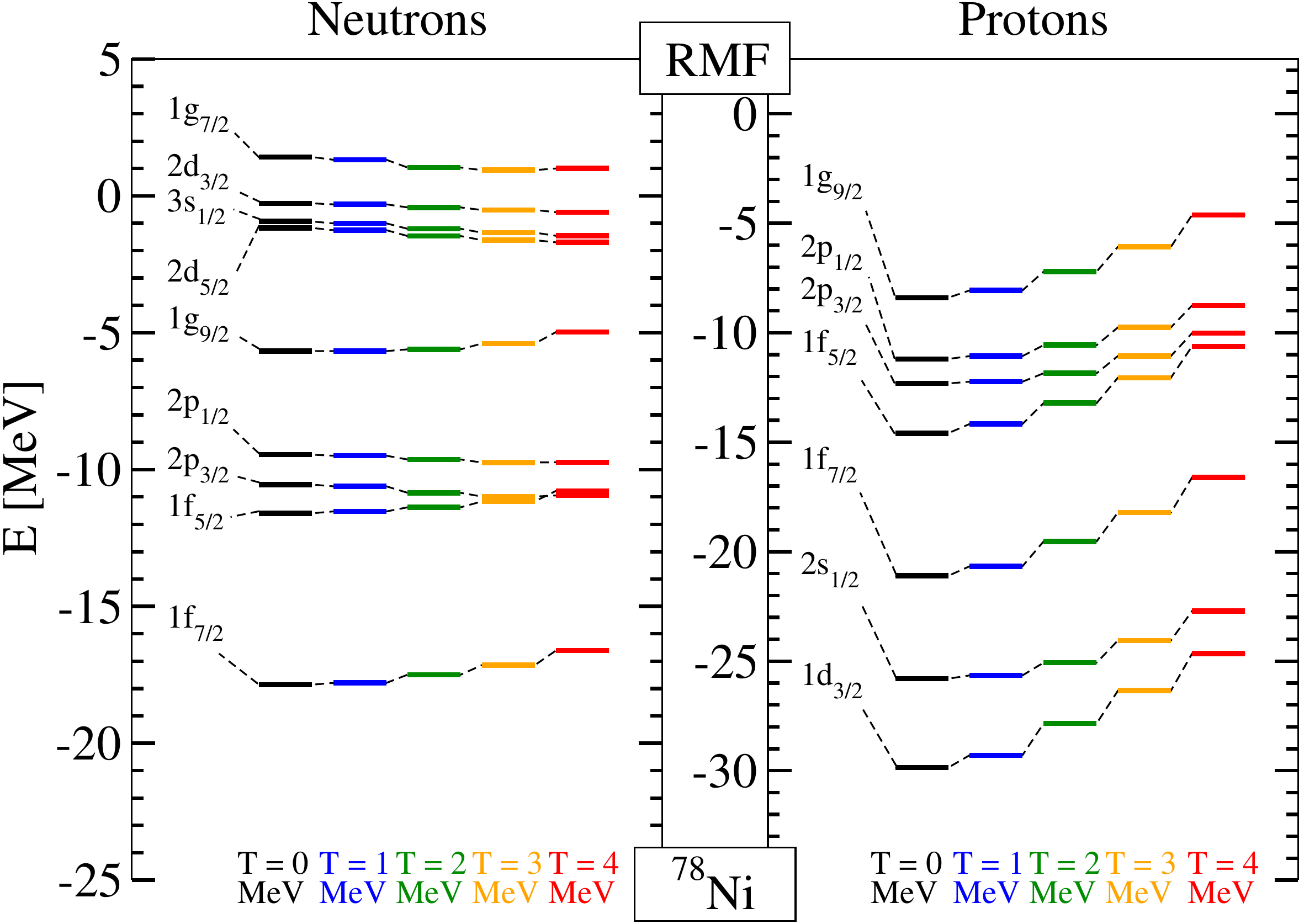}
\caption{} 
\label{Ni78_RMF} 
\end{subfigure}
\caption{Single-particle states in (a) $^{68}\text{Ni}$, (b) $^{70}\text{Ni}$, (c) $^{72}\text{Ni}$, (d) $^{74}\text{Ni}$, (e) $^{76}\text{Ni}$, and (f) $^{78}\text{Ni}$ isotopes at zero and finite temperature calculated within the RMF approximation.}
\end{figure*}

The reference single-particle spectra  for $^{68-78}\text{Ni}$ isotopes obtained within the RMF(+BCS) approach at zero and finite temperature are shown in Figs. \ref{Ni68_RMF}-\ref{Ni78_RMF}. At $T=0$, the sizable effect of neutron pairing correlations is restricted by the neighborhood of the Fermi surface. This is reflected in the presented spectra, while the pure single-particle levels computed without pairing at $T = 0$ are not very different from those at $T = 1$ MeV. As a result of pairing correlations, the neutron states above (below) Fermi surface are displaced towards higher (lower) energies. Since the chemical potential of the neutron subsystem increases with the increase of the neutron number, the intruder state $1\text{g}_{9/2}$, shifted upward in $^{68-72}\text{Ni}$ isotopes, is  shifted downward in $^{74-76}\text{Ni}$ isotopes. 
As it is mentioned above, superfluid pairing does not show up in the closed-shell $^{78}\text{Ni}$ and in the closed-shell proton subsystems of $^{68-76}\text{Ni}$ isotopes, although the proton states are implicitly affected by the neutron superfluidity via the self-consistent mean field.

In general, one observes that the neutron addition induces the displacement of the proton states altogether towards lower energies. As a result, the two proton shell gaps, which are associated with the magic numbers 20 and 28, are slightly diminished with the increase of the neutron number. In the neutron subsystems of $^{68-76}\text{Ni}$ isotopes, one observes an abrupt change of the neutron mean-field energies at the temperature between $T=0$ and $T=1$ MeV due to the superfluid phase transition at the critical temperatures $T_{c}$, given in Table \ref{Neutron pairing gaps and critical T}. In contrast, the doubly-magic $^{78}\text{Ni}$ nucleus shows a smooth development of the neutron mean-field energies due to the absence of the superfluid phase transition. As the temperature increases from $T=1$ MeV to $T=4$ MeV, the neutron mean-field states in the major shell display a tendency to densifying.  Analogously to the case of $T=0$, at $T>0$ the proton mean-field states of  isotopes with larger neutron numbers also have lower energies. The overall trend of the proton mean-field energies exhibits a gradual increment of 1-2 MeV in the $1\leq T\leq 4$ MeV temperature interval. 


\begin{figure*}[h!]
\begin{subfigure}{0.4\textwidth}
\centering
\includegraphics[scale=0.35]{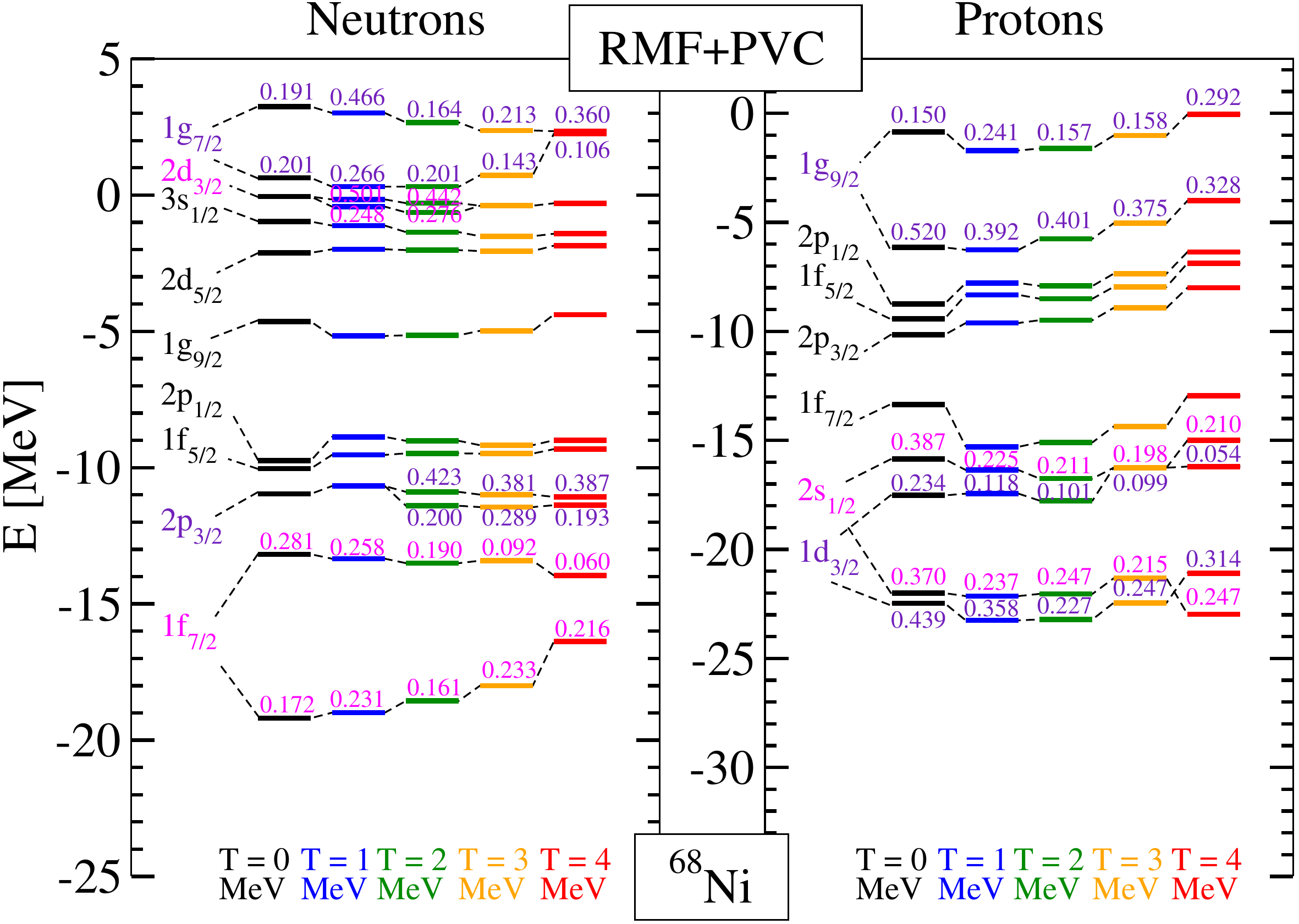} 
\caption{}
\label{Ni68_RMF+PVC} 
\end{subfigure}
\hfill
\begin{subfigure}{0.5\textwidth}
\centering
\includegraphics[scale=0.35]{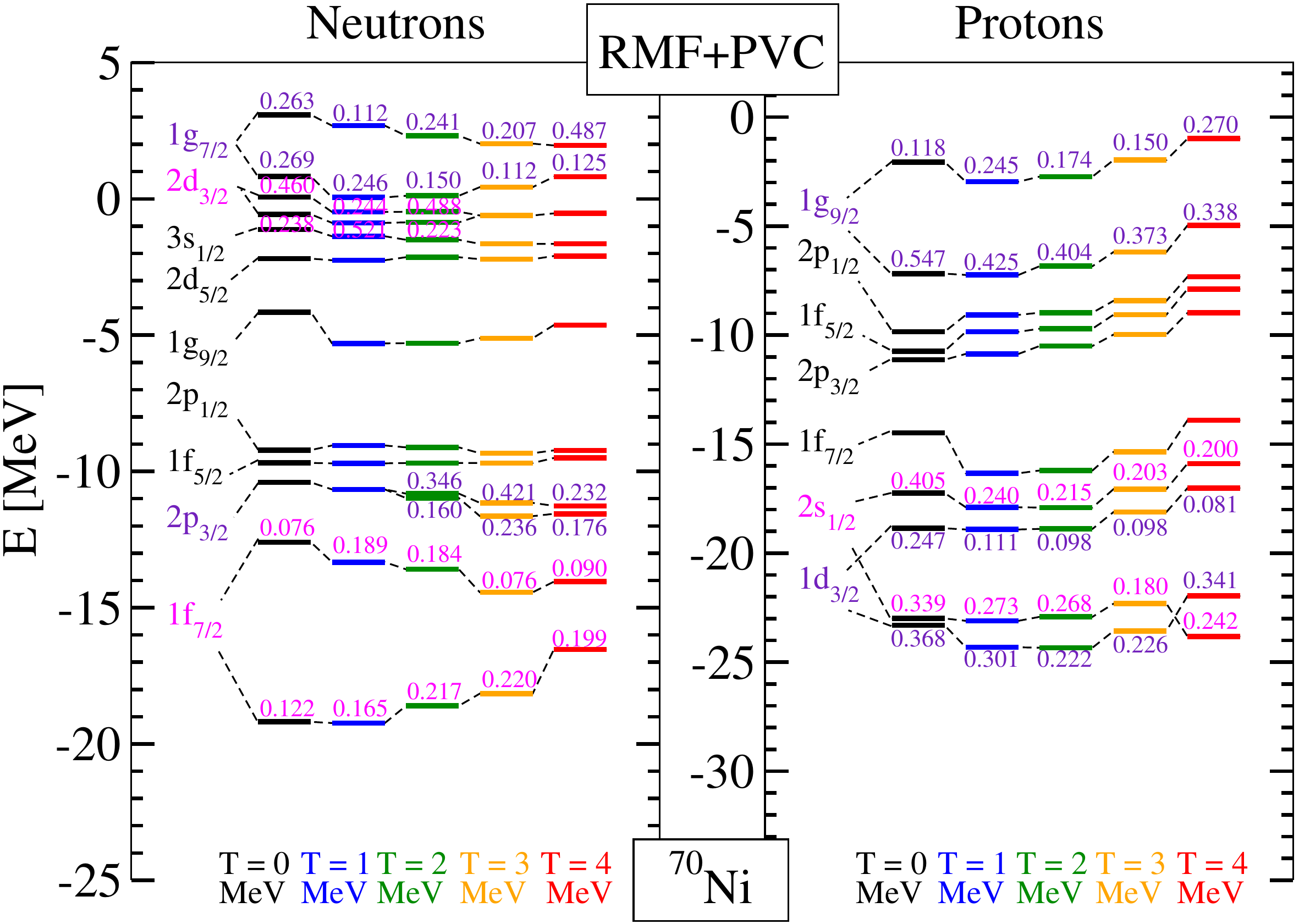}
\caption{}
\label{Ni70_RMF+PVC}
\end{subfigure}
\hfill
\begin{subfigure}{0.4\textwidth}
\centering
\includegraphics[scale=0.35]{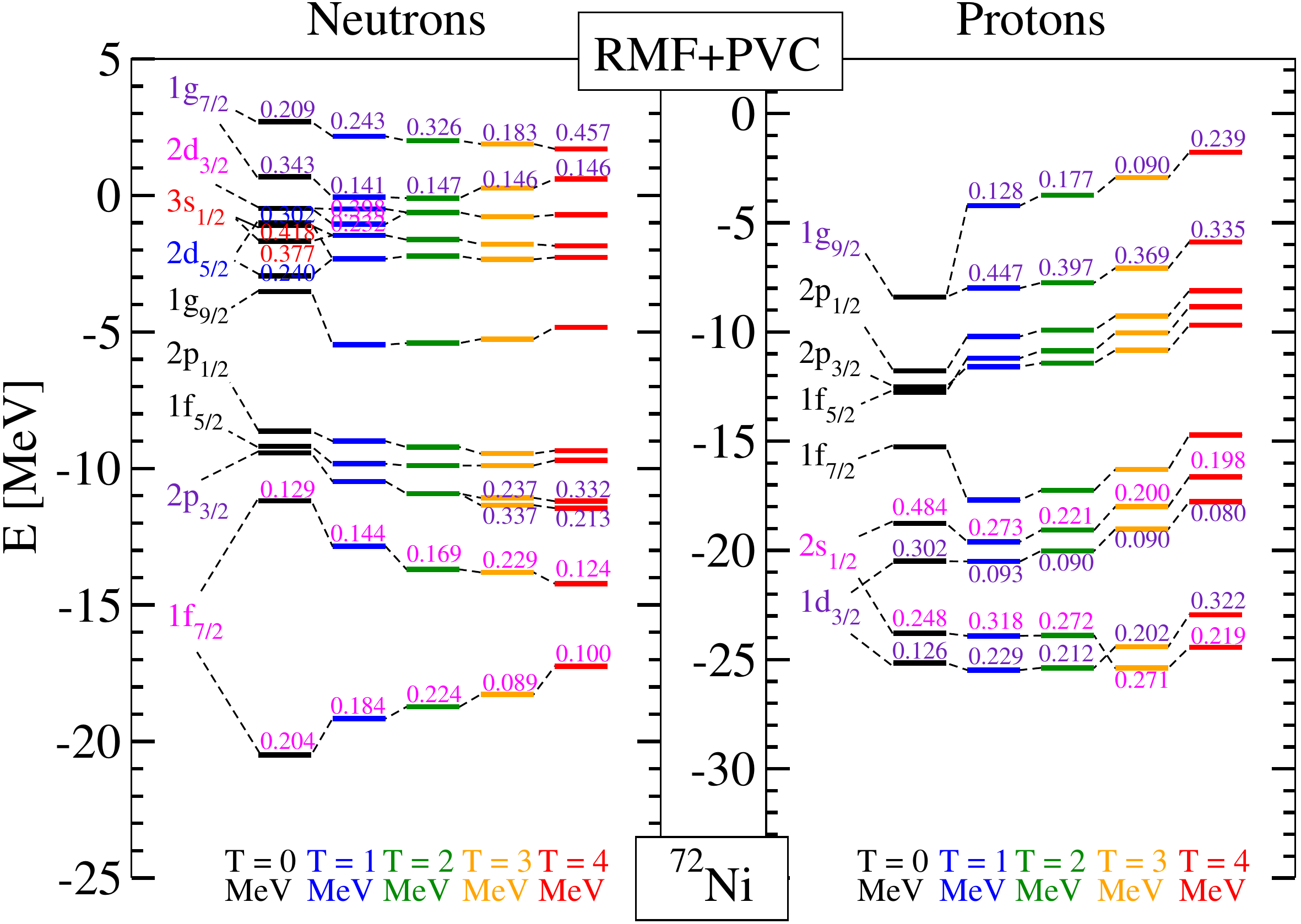}
\caption{}
\label{Ni72_RMF+PVC} 
\end{subfigure}
\hfill
\begin{subfigure}{0.5\textwidth}
\centering
\includegraphics[scale=0.35]{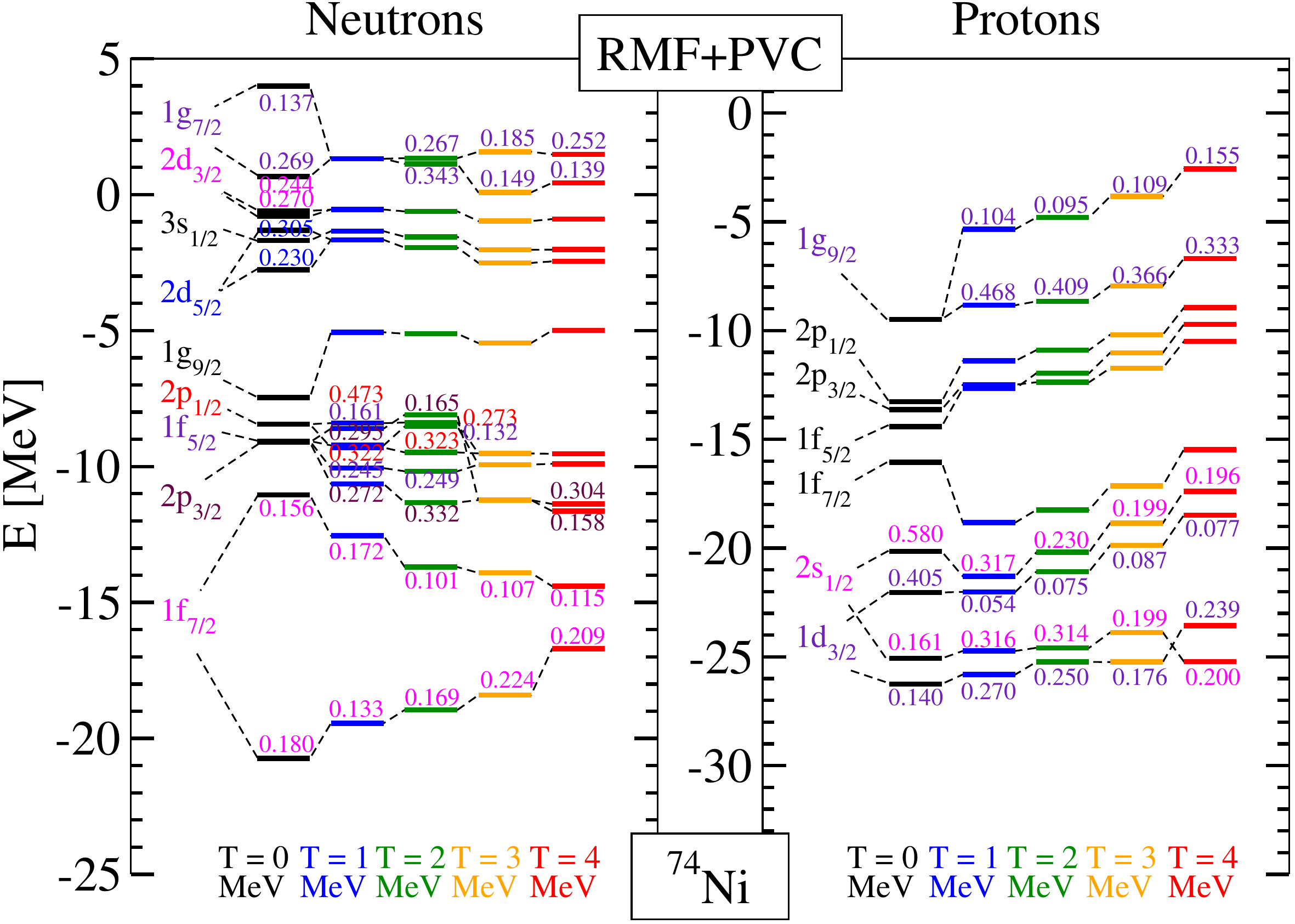}
\caption{} 
\label{Ni74_RMF+PVC} 
\end{subfigure}
\hfill
\begin{subfigure}{0.4\textwidth}
\centering
\includegraphics[scale=0.35]{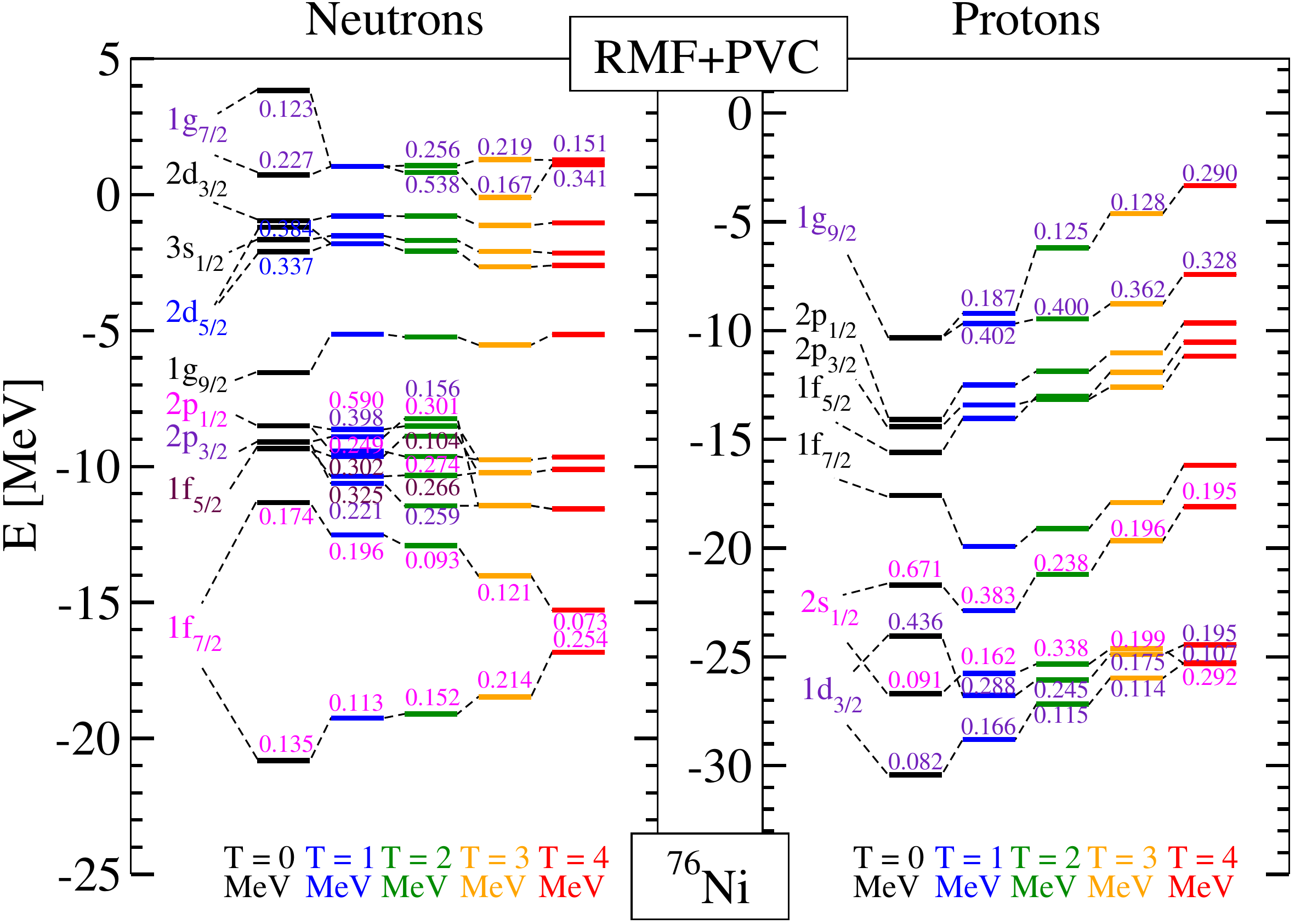}
\caption{} 
\label{Ni76_RMF+PVC} 
\end{subfigure}
\hfill
\begin{subfigure}{0.5\textwidth}
\centering
\includegraphics[scale=0.35]{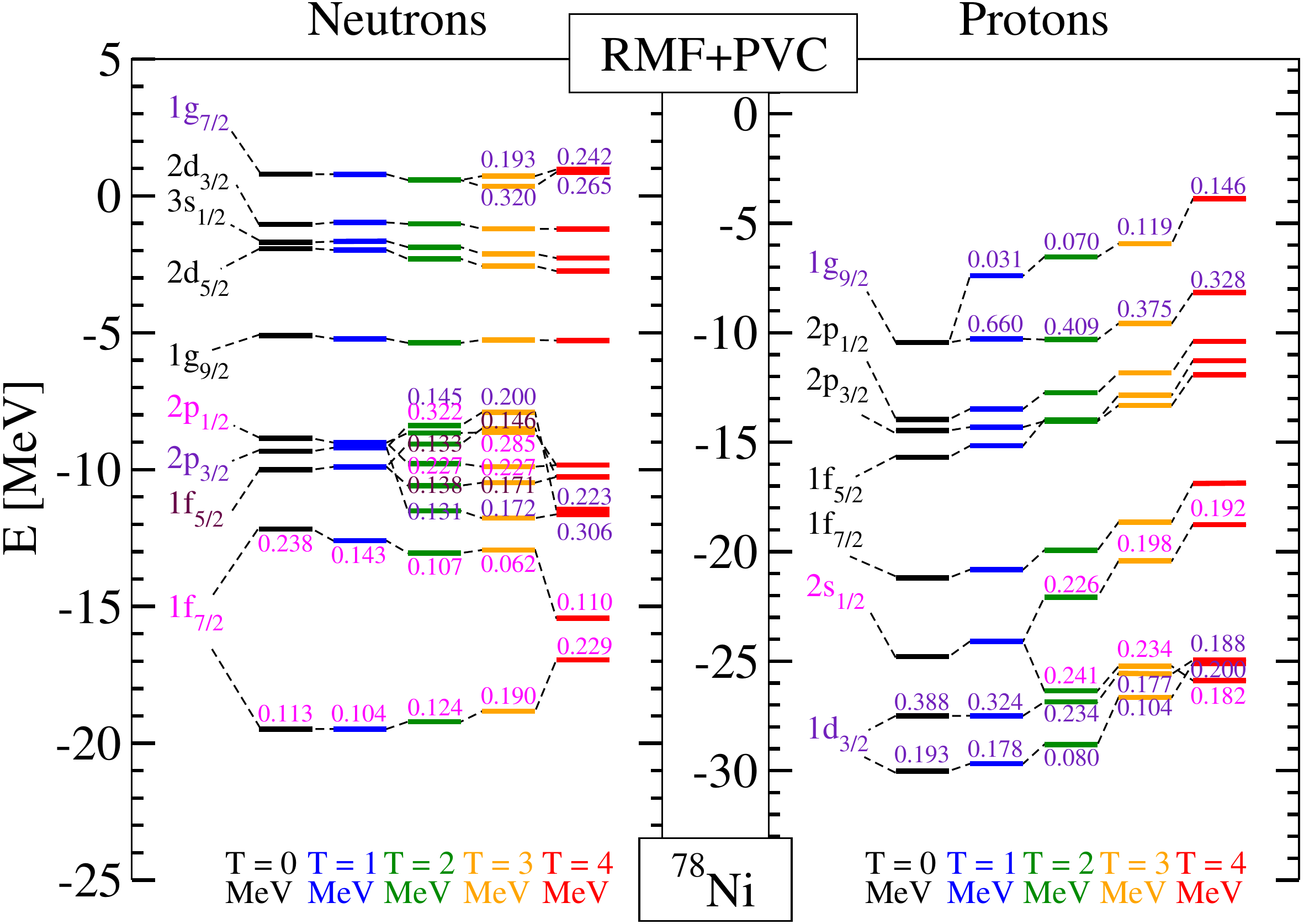}
\caption{} 
\label{Ni78_RMF+PVC} 
\end{subfigure}
\caption{The dominant fragments of the single-particle states in (a) $^{68}\text{Ni}$, (b) $^{70}\text{Ni}$, (c) $^{72}\text{Ni}$, (d) $^{74}\text{Ni}$, (e) $^{76}\text{Ni}$, and (f) $^{78}\text{Ni}$ isotopes at zero and finite temperature calculated in the RMF+PVC approximation. The labels of strongly fragmented states, together with their shared spectroscopic factors, are written in color. The labels of good single-particle states and weakly fragmented states are written in black and represented by a single dominant fragment.}
\end{figure*}

Figs. \ref{Ni68_RMF+PVC}-\ref{Ni78_RMF+PVC} display the dominant fragments of both neutron and proton single-particle states for $^{68-78}\text{Ni}$ isotopes computed within the RMF+PVC approach. As in Refs. \cite{Litvinova2006a, Litvinova2012, Wibowo2020}, we select the dominant fragment for each mean-field state according to the spectroscopic factors of the fragments. Generally, the dominant fragment is the one with the largest spectroscopic factor. In the vicinity of the Fermi energy there is typically one fragment with the spectroscopic factor of $\sim 0.7-0.9$, while the other fragments are characterized by considerably smaller spectroscopic factors (below 0.1). This fragmentation pattern is preserved at all temperatures. The neutron $1\text{g}_{9/2}$  state and the proton $1\text{f}_{7/2}$ state of $^{68-78}\text{Ni}$ isotopes are the examples of such pattern and called good single-particle states. The dominant fragments of these states have the energies rather close to the energies of the original mean-field states. 
The states far away from the Fermi surface are either strongly or weakly fragmented. The strongly fragmented states are characterized by the presence of two competing fragments with comparable spectroscopic factors. The first fragment is chosen to be the fragment with the largest spectroscopic factor, while another fragment has the spectroscopic factor not smaller than 40\% of the spectroscopic factor of the first fragment. Analogously to the good single-particle states, these two dominant fragments have energies close to their original mean-field energies. 
The weakly fragmented states are characterized by one fragment with a dominant spectroscopic factor, while other fragments have spectroscopic factors smaller than 40\% of the spectroscopic factor of the first fragment. In this case, the dominant fragment is well defined as the one with the largest spectroscopic factor. The energy of the dominant fragment is again close to the original mean-field energy. 
The degree of fragmentation of a state far away from Fermi surface can vary with temperature. One example is the neutron $2\text{p}_{3/2}$ state in $^{74}\text{Ni}$, as displayed in Fig. \ref{Ni74_RMF+PVC}. While the neutron $2\text{p}_{3/2}$ state is weakly fragmented at $T=0$ and $T=3$ MeV, it is strongly fragmented at $T=1$ MeV, $T=2$ MeV, and $T=4$ MeV with 0.30/0.27, 0.17/0.33, and 0.30/0.16 share of spectroscopic factors, respectively. The degree of fragmentation of the same state can also vary along the isotopic chain. At $T=1$, the neutron $2\text{p}_{3/2}$ state is a good single-particle state in $^{68,70,72}\text{Ni}$ (see Figs. \ref{Ni68_RMF+PVC}-\ref{Ni72_RMF+PVC}), whereas it is a strongly fragmented state in $^{74,76}\text{Ni}$ (see Figs. \ref{Ni74_RMF+PVC} and \ref{Ni76_RMF+PVC}). At the same temperature, the neutron $2\text{p}_{3/2}$ state is a weakly fragmented state in the doubly-magic $^{78}\text{Ni}$ nucleus (Fig. \ref{Ni78_RMF+PVC}). In contrast to the neutron $2\text{p}_{3/2}$ state, which can be strongly fragmented at some temperatures, the proton $2\text{p}_{1/2}$, $1\text{f}_{5/2}$, and $2\text{p}_{3/2}$ states of $^{68-78}\text{Ni}$ isotopes are either good single-particle states or weakly fragmented states, as shown in the right panels of Figs. \ref{Ni68_RMF+PVC}-\ref{Ni78_RMF+PVC}.

\begin{figure}
\includegraphics[scale=0.35]{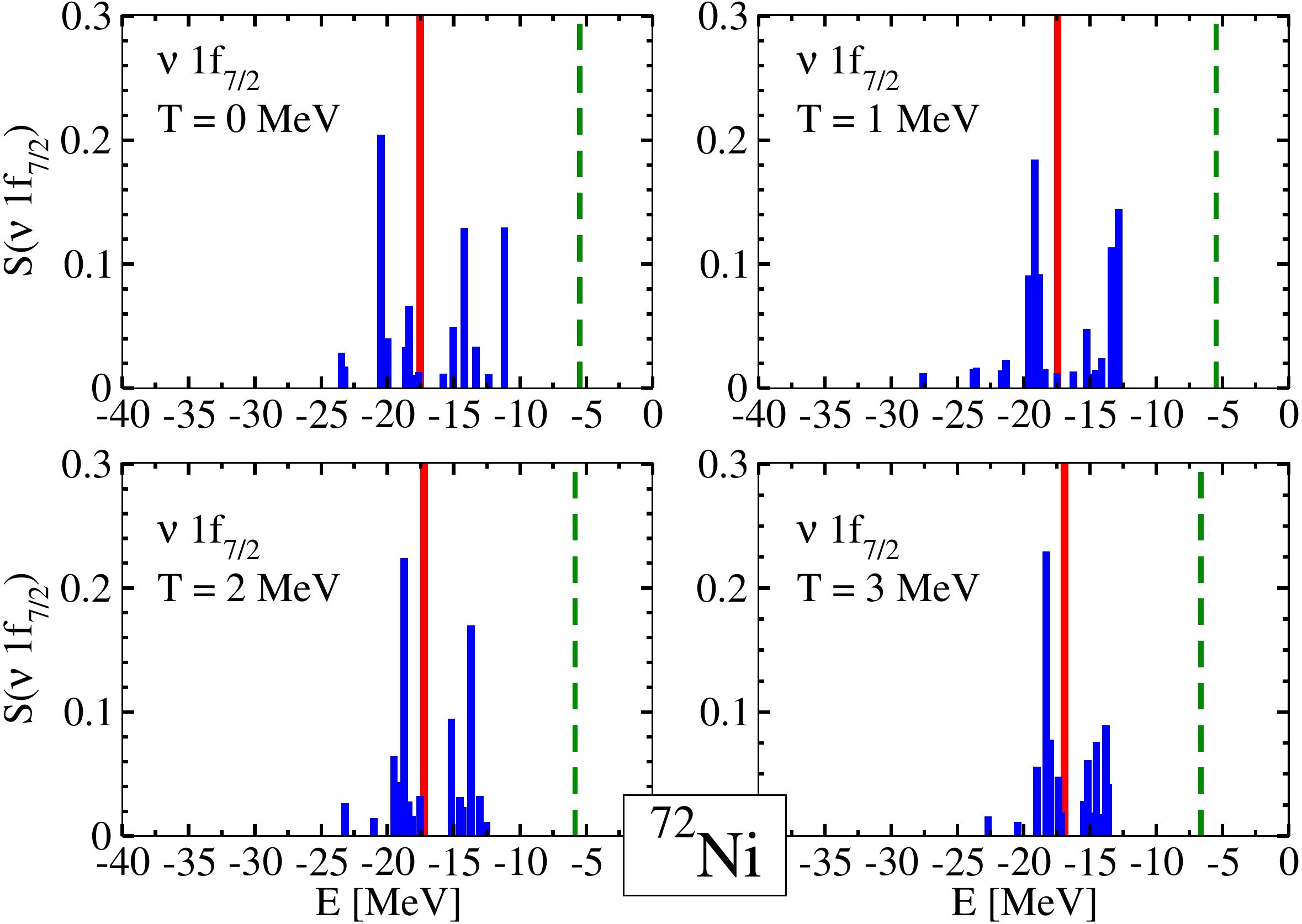}
\caption{Temperature evolution of the neutron $1\text{f}_{7/2}$ state of $^{72}\text{Ni}$ isotope. The pure mean-field state is represented by the red bar, while the blue bars indicate the spectroscopic factors of the fragmented states. The green line corresponds to the chemical potential.} 
\label{Ni72_n_1f7over2} 
\end{figure}

\begin{figure}
\includegraphics[scale=0.35]{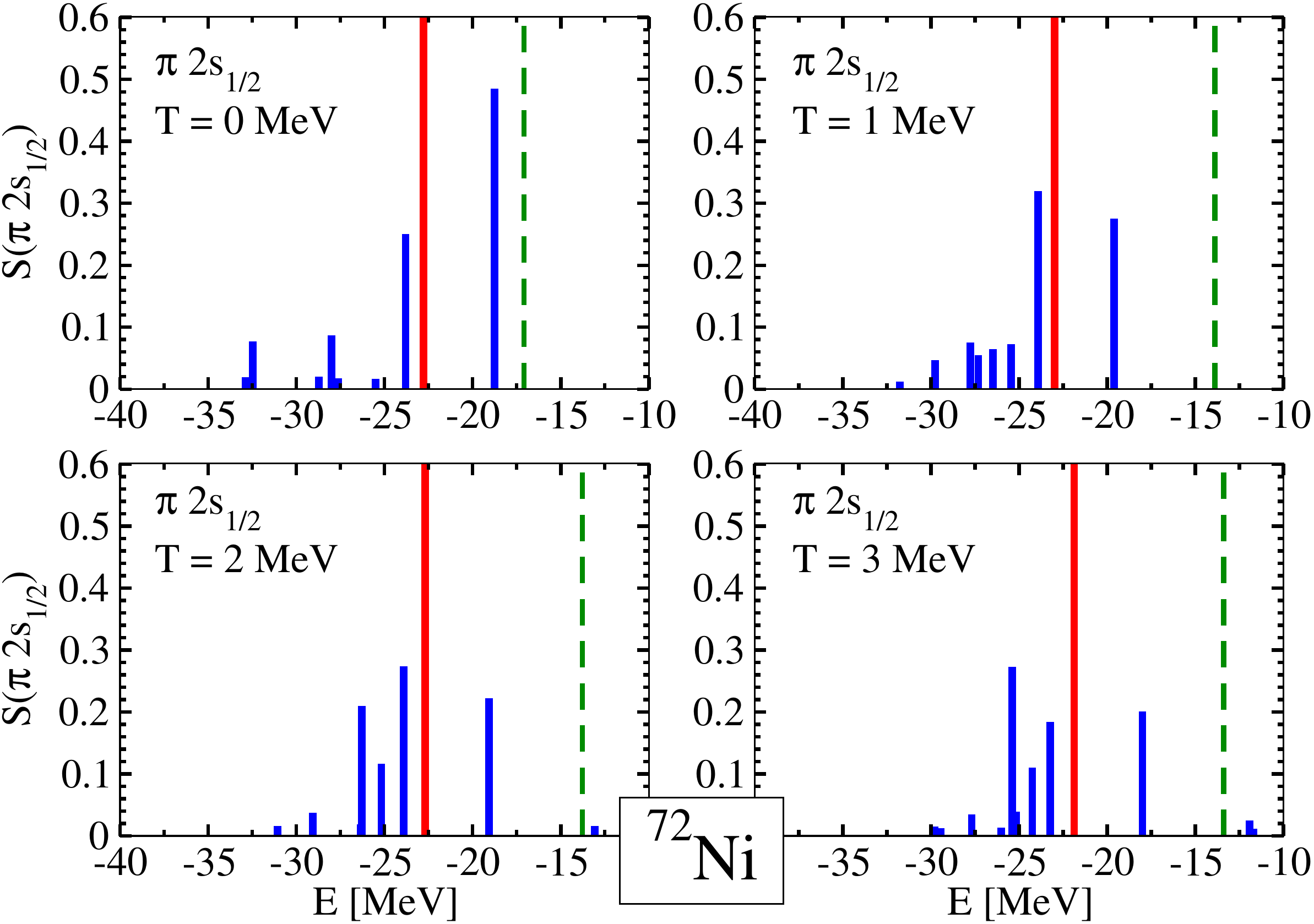}
\caption{\label{Ni72_p_2s1over2} Same as Fig. \ref{Ni72_n_1f7over2}, but for proton $2\text{s}_{1/2}$ state of $^{72}\text{Ni}$ isotope.} 
\end{figure}

For the states remote from the Fermi surface, one often encounters two or more fragments, which exhibit comparable spectroscopic factors. An example of such states is the neutron $1\text{f}_{7/2}$ state of $^{68-78}\text{Ni}$ isotopes. The temperature evolution of this state in $^{72}\text{Ni}$ is illustrated in Fig. \ref{Ni72_n_1f7over2}. In general, we observe a consistent fragmentation pattern, which is preserved throughout all temperatures. For each temperature, there exists a cluster of fragments with the energies lower than the mean-field energy (low-energy cluster), and another cluster with the energies larger than the mean-field energy (high-energy cluster). Each cluster has one or two major fragments with relatively large spectroscopic factors. In this case, the major fragments play a role of the dominant fragments. At $T=0$, the dominant fragments consist of three major fragments. The low-energy fragment has the spectroscopic factor 0.20, while each of the other two fragments has the spectroscopic factor 0.13. The phase transition, which occurs around $T=1$ MeV, together with the beginning of the thermal unblocking, modifies the strength distribution from a predominantly three-peak structure to a predominantly two-peak structure. As the temperature increases from $T=2$ MeV to $T=4$ MeV, the low-energy major fragment dominates, and the high-energy cluster exhibits a strong fragmentation. 
For comparison, the temperature evolution of the proton $2\text{s}_{1/2}$ state is illustrated in Fig. \ref{Ni72_p_2s1over2}. Analogously to the neutron $1\text{f}_{7/2}$ state, the proton $2\text{s}_{1/2}$ state also exhibits a two-cluster structure, where the clusters are located on the opposite sides of the original mean-field state. At $T=0$, the dominant fragments are the two major fragments with 0.25/0.48 share of the spectroscopic factors. As the temperature increases, the low-energy cluster becomes strongly fragmented, while the high-energy major fragment in the vicinity of the Fermi energy behaves as a good single-particle state. A similar fragmentation pattern holds for the proton $2\text{s}_{1/2}$ state in other Ni isotopes.
Another example from the neutron subsystem is represented by the neutron $1\text{g}_{7/2}$ state, which is a particle state, i.e., located above the Fermi energy. In $^{68-72}\text{Ni}$ isotopes, this state exhibits the fragmentation pattern, which is similar to that of the neutron $1\text{f}_{7/2}$ state in all temperature regimes, as shown in Figs. \ref{Ni68_RMF+PVC}-\ref{Ni72_RMF+PVC}. However, in $^{74-78}\text{Ni}$ isotopes, the state $1\text{g}_{7/2}$ suddenly becomes a good single-particle state at $T=1$ MeV, though it is gradually becoming more fragmented at higher temperatures (see Figs. \ref{Ni74_RMF+PVC}-\ref{Ni78_RMF+PVC}). 

Thereby, while the evolution of the good single-particle states in all Ni isotopes is quite similar to the evolution of the corresponding mean-field states, the evolution of the states remote from the Fermi surface exhibits various scenarios in both neutron and proton subsystems. However, the evolution of single-particle states in the proton subsystem remains almost unchanged across the Ni isotopic chain. In contrast, a significant modification of the fragmentation pattern occurs in the neutron subsystem. Therefore, we confine our further discussion by only the neutron single-particle states. We examine some simplistic (toy) models of varying complexity below to qualitatively describe the essential factors that determine the fragmentation patterns of single-particle states.




\subsection{Lessons from the toy models\label{Toy models}}

Let us first consider a system which consists of one state $k_{1}$ and one phonon with frequency $\omega_{1}$. Under these conditions,  Eq. \eqref{Sigma e} for the diagonal mass operator $\Sigma^{e}_{k_{1}}(\varepsilon)$ takes the form: 
\begin{eqnarray}
\Sigma^{e}_{k_{1}}(\varepsilon)&=g^{1\ast}_{k_{1}k_{1}}g^{1}_{k_{1}k_{1}}&\left\{\frac{N(\omega_{1},T)+1-n(\varepsilon_{k_{1}}-\mu,T)}{\varepsilon-\varepsilon_{k_{1}}+\mu-\omega_{1}}\right.\nonumber\\
&&+\left.\frac{n(\varepsilon_{k_{1}}-\mu,T)+N(\omega_{1},T)}{\varepsilon-\varepsilon_{k_{1}}+\mu+\omega_{1}}\right\}.
\end{eqnarray}
The diagonalization of the matrix
\begin{equation}
\left( \begin{array}{ccc}
\varepsilon_{k_{1}}-\mu & \xi^{1(+1)}_{k_{1}k_{1}} & \xi^{1(-1)}_{k_{1}k_{1}} \\ 
\xi^{1(+1)\ast}_{k_{1}k_{1}} & \varepsilon_{k_{1}}-\mu-\omega_{1} & 0 \\ 
\xi^{1(-1)\ast}_{k_{1}k_{1}} & 0 & \varepsilon_{k_{1}}-\mu+\omega_{1} \\ 
\end{array}  \right),
\end{equation}
where
\begin{equation}
\xi^{1(\pm 1)}_{k_{1}k_{1}}=g^{1(\pm 1)}_{k_{1}k_{1}}\sqrt{N(\omega_{1},T)+n(\pm(\varepsilon_{k_{1}}-\mu),T)},
\end{equation}
results in three different energies $\varepsilon_{k_{1}}^{(\lambda)}$ ($\lambda=1,\;2,\;3$). For each $\lambda$, the corresponding spectroscopic factor $S^{(\lambda)}_{k_{1}}$ reads 
\begin{eqnarray}
\label{S1state1phonon}S_{k_{1}}^{(\lambda)}&=\Bigg\{1+g^{1\ast}_{k_{1}k_{1}}&g^{1}_{k_{1}k_{1}}\Bigg[\frac{N(\omega_{1},T)+1-n(\varepsilon_{k_{1}}-\mu,T)}{[\varepsilon-\varepsilon_{k_{1}}+\mu-\omega_{1}]^{2}}\nonumber\\
&&+\frac{n(\varepsilon_{k_{1}}-\mu,T)+N(\omega_{1},T)}{[\varepsilon-\varepsilon_{k_{1}}+\mu+\omega_{1}]^{2}}\Bigg]\Bigg\}^{-1}_{\varepsilon=\varepsilon_{k_{1}}^{(\lambda)}}.
\end{eqnarray}

\begin{figure}[h]
\includegraphics[scale=0.45]{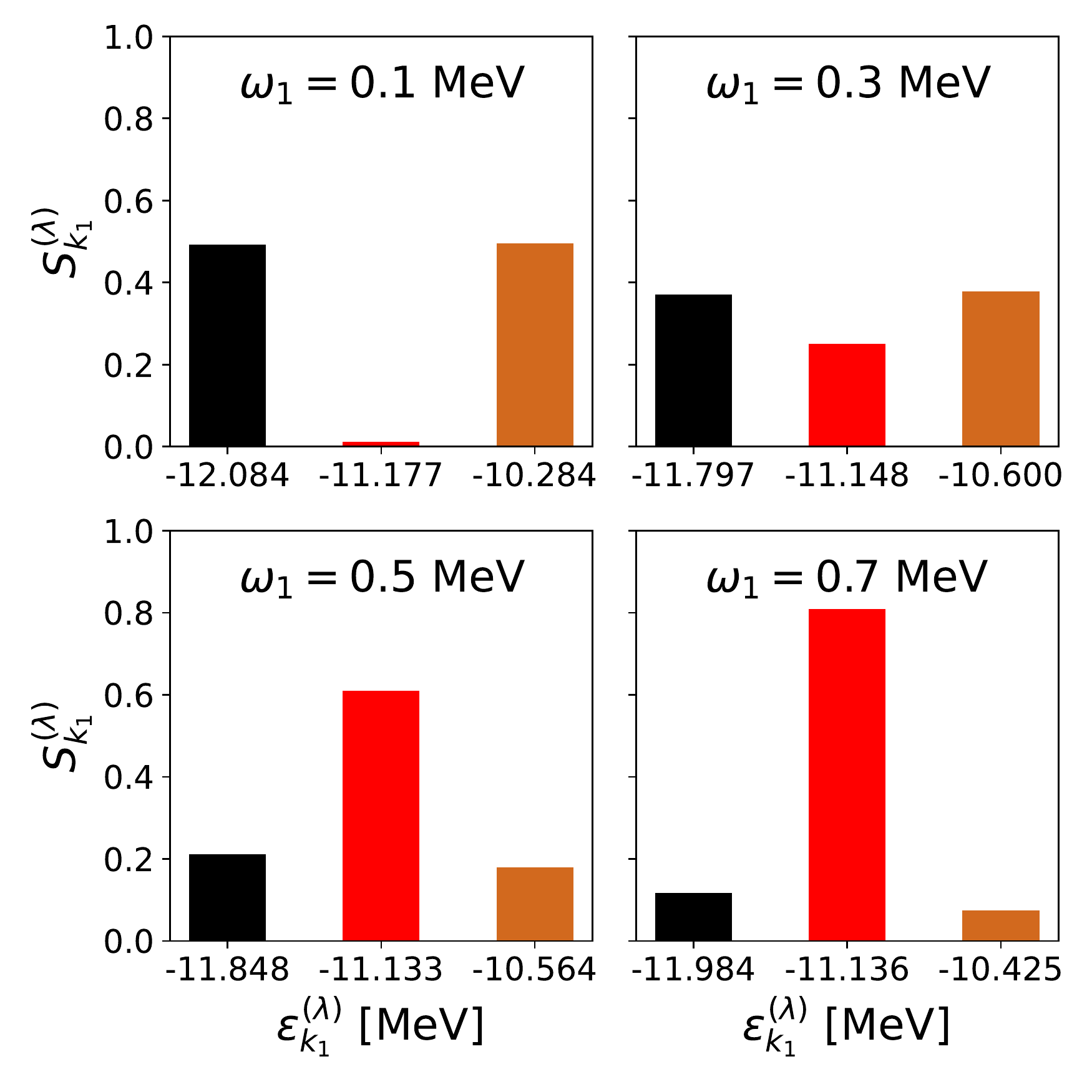}
\caption{\label{3-fragment}The evolution of spectroscopic factors $S^{(\lambda)}_{k_{1}}$ for the state $1\text{f}_{7/2}$ with the phonon frequency $\omega_{1}$ at $T=1$ MeV. The colors black, red, and chocolate correspond to $\varepsilon_{k_{1}}^{(1)}$, $\varepsilon_{k_{1}}^{(2)}$, and $\varepsilon_{k_{1}}^{(3)}$, respectively.} 
\end{figure} 

\begin{figure}[h]
\includegraphics[scale=0.50]{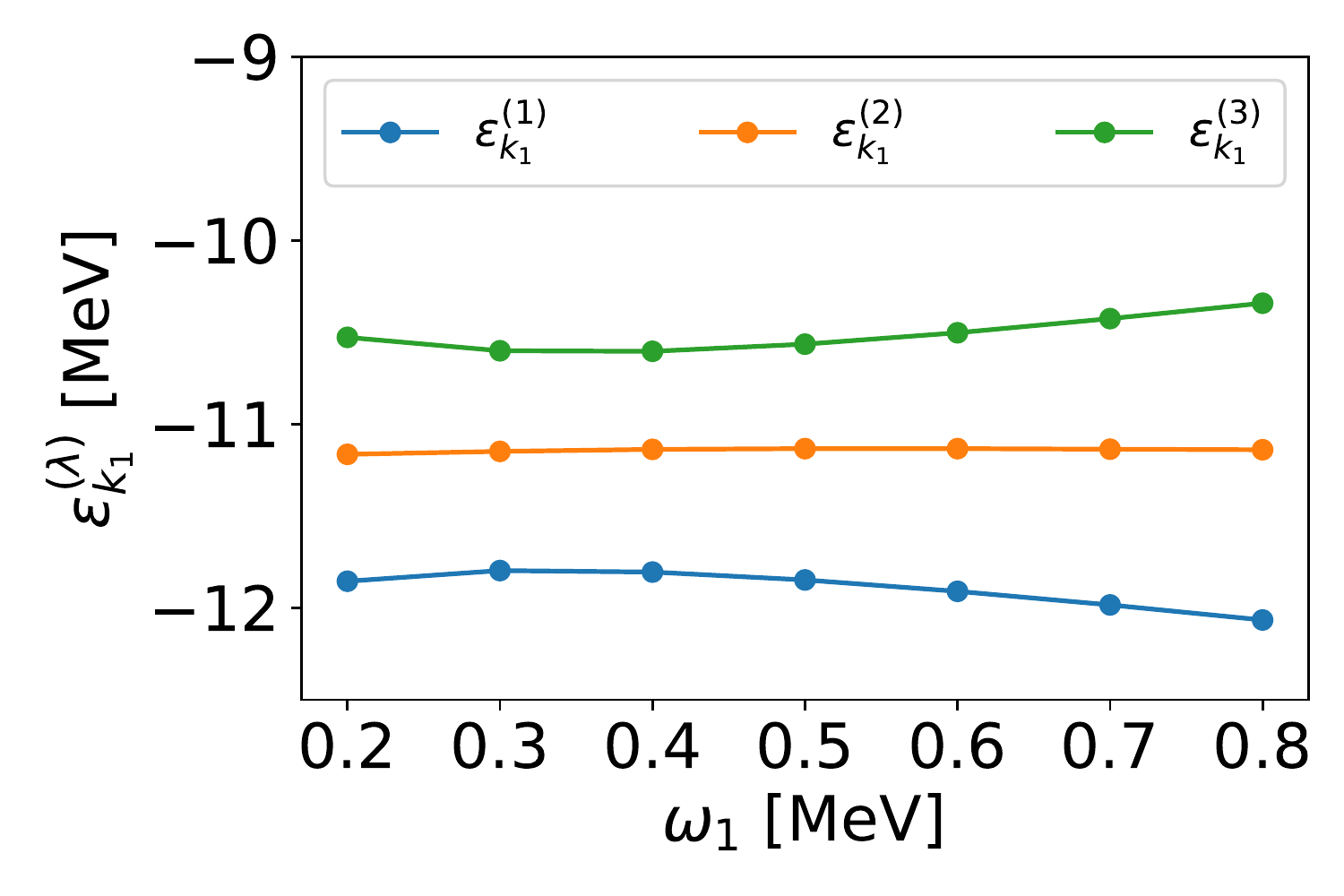}
\caption{\label{DeltaE of 3-fragment} The evolution of the energy fragments $\varepsilon_{k_{1}}^{(\lambda)}$ with the phonon frequency $\omega_{1}$ at $T=1$ MeV.} 
\end{figure} 

\begin{figure}[h]
\includegraphics[scale=0.40]{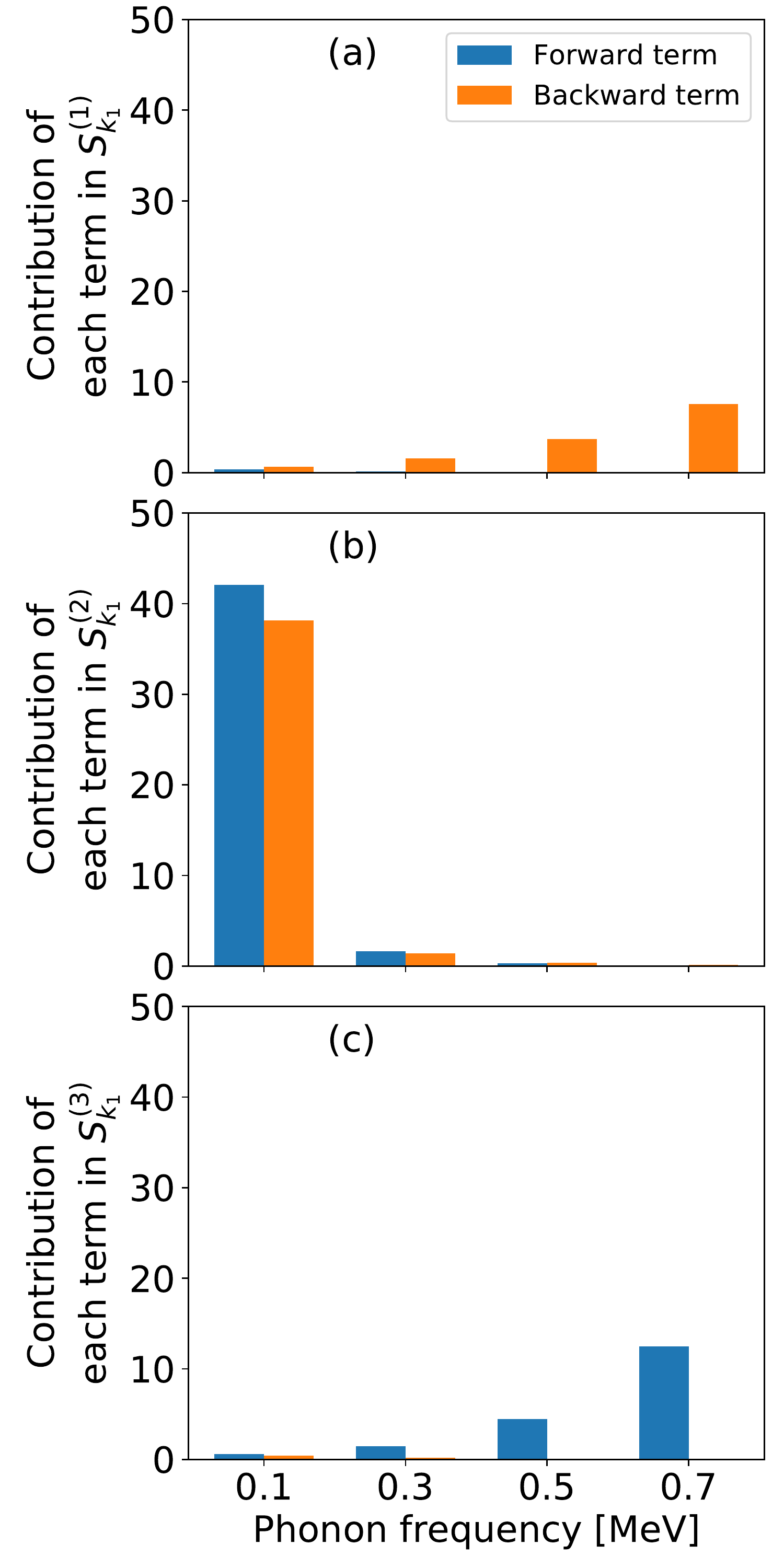}
\caption{\label{TermbyTerm_3-fragment}The interplay between the forward and backward going terms of the spectroscopic factors $S^{(1,2,3)}_{k_{1}}$ for the energy fragments (a) $\varepsilon_{k_{1}}^{(1)}$, (b) $\varepsilon_{k_{1}}^{(2)}$, and (c) $\varepsilon_{k_{1}}^{(3)}$.} 
\end{figure} 

\begin{figure}[h]
\includegraphics[scale=0.45]{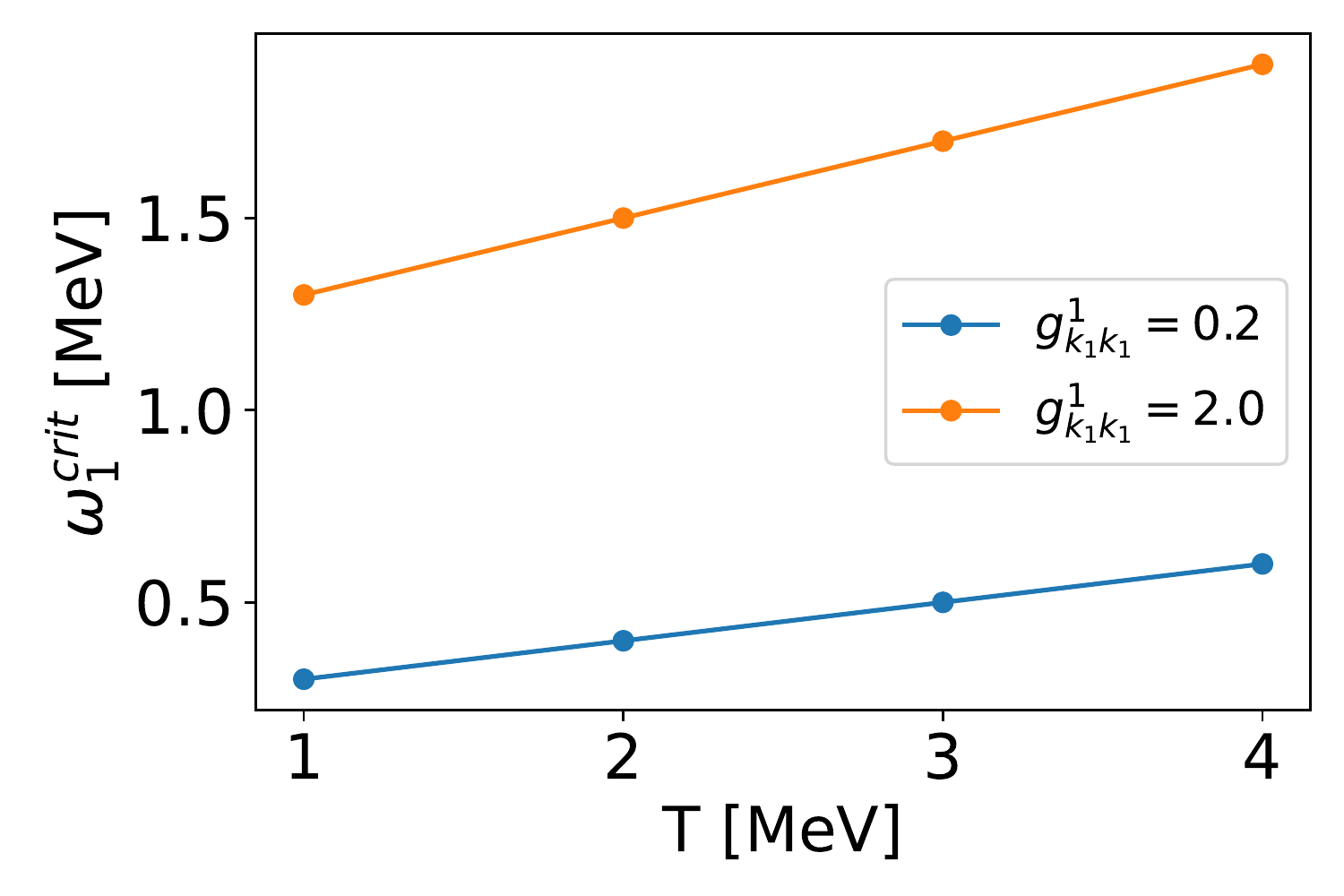}
\caption{\label{Critical_omega}The evolution of the critical phonon frequency $\omega_{1}^{\text{crit}}$ with temperature $T$ for $g^{1}_{k_{1}k_{1}}=0.2$ and $g^{1}_{k_{1}k_{1}}=2.0$.} 
\end{figure} 

To illustrate this model, we computed the spectroscopic factors $S^{(\lambda)}_{k_{1}}$ for the specific neutron state $1\text{f}_{7/2}$ of $^{70}\text{Ni}$ nucleus at fixed temperature $T=1$ MeV. From the thermal RMF calculations, the RMF energy $\varepsilon_{k_{1}}$ of the state $1\text{f}_{7/2}$ and the chemical potential $\mu$ at $T=1$ MeV are obtained as $-17.296$ MeV and $-6.114$ MeV, respectively. The phonon vertex $g^{1}_{k_{1}k_{1}}$ is taken equal to 0.2, which is a typical value for the phonon vertices calculated within FT-RRPA. Fig. \ref{3-fragment} demonstrates the evolution of the spectroscopic factors $S_{k_{1}}^{(\lambda)}$ with the phonon frequency $\omega_{1}$. As one can see from Fig. \ref{3-fragment}, both first and third energy fragments are dominant at $\omega_{1}<0.3$ MeV, while the second fragment becomes the dominant state at $\omega_{1}>0.3$ MeV. The critical phonon frequency $\omega^{\text{crit}}_{1}=0.3$ MeV refers to the phonon frequency, where the second fragment starts becoming dominant, and the number of the competing spectroscopic factors is maximal, i.e., 3. A high number of the competing spectroscopic factors  
can be associated with strong PVC.
As shown in Fig. \ref{DeltaE of 3-fragment}, the energy differences $\varepsilon_{k_{1}}^{(2)}-\varepsilon_{k_{1}}^{(1)}$ and $\varepsilon_{k_{1}}^{(3)}-\varepsilon_{k_{1}}^{(2)}$ are minimal at $\omega_{1}=\omega_{1}^{\text{crit}}$. This implies 
a correlation between the proximity of the fragments to each other and the degree of fragmentation.
Eq. \eqref{S1state1phonon} suggests that the evolution of the spectroscopic factors $S_{k_{1}}^{(\lambda)}$ with the phonon frequency $\omega_{1}$ is determined by the interplay between the forward going term (the first term in the square bracket) and the backward going term (the second term in the square bracket). To better understand this interplay, we track the evolution of the forward and the backward going terms with the phonon frequency $\omega_{1}$ for each energy fragment $\varepsilon_{k_{1}}^{(\lambda)}$, as displayed in Fig. \ref{TermbyTerm_3-fragment}. As shown in Fig. \ref{TermbyTerm_3-fragment}b, there is an almost equal contribution of the forward and the backward going terms to the spectroscopic factor $S^{(2)}_{k_{1}}$ for all phonon frequencies $\omega_{1}$, while the backward(forward) going term is always dominant in the first (third) energy fragment, as shown in Figs. \ref{TermbyTerm_3-fragment}a  (\ref{TermbyTerm_3-fragment}c). At $\omega_{1}<0.3$ MeV, the total contribution of the forward and the backward going terms to the second energy fragment is higher than to the other two fragments, leading to the higher spectroscopic factors.
Starting from $\omega^{\text{crit}}_{1}=0.3$ MeV, the 
rapid growth of the backward (forward) going terms in the inverse spectroscopic factor $S^{(1)}_{k_{1}}$ ($S^{(3)}_{k_{1}}$) is associated with a quick increase of the spectroscopic factor $S^{(2)}_{k_{1}}$ of the second fragment. Similar trends are observed for different temperatures $T$ and different values of the phonon vertex $g^{1}_{k_{1}k_{1}}$. For a fixed phonon vertex $g^{1}_{k_{1}k_{1}}=0.2$, one observes an increase of the critical phonon frequency $\omega_{1}^{\text{crit}}$ by the amount of roughly 0.1 MeV per 1 MeV temperature step. However, a somewhat faster increase of $\omega_{1}^{\text{crit}}$ is recorded when one increases the value of $g^{1}_{k_{1}k_{1}}$ by one order of magnitude, as seen in Fig. \ref{Critical_omega}. Thus, the first toy model demonstrates how the evolution of the spectroscopic factors $S_{k_{1}}^{(\lambda)}$ with temperature $T$ is governed by the phonon frequency $\omega_{1}$ and the magnitude of the phonon vertex $g_{k_{1}k_{1}}^{1}$.

\begin{figure}[h]
\includegraphics[scale=0.20]{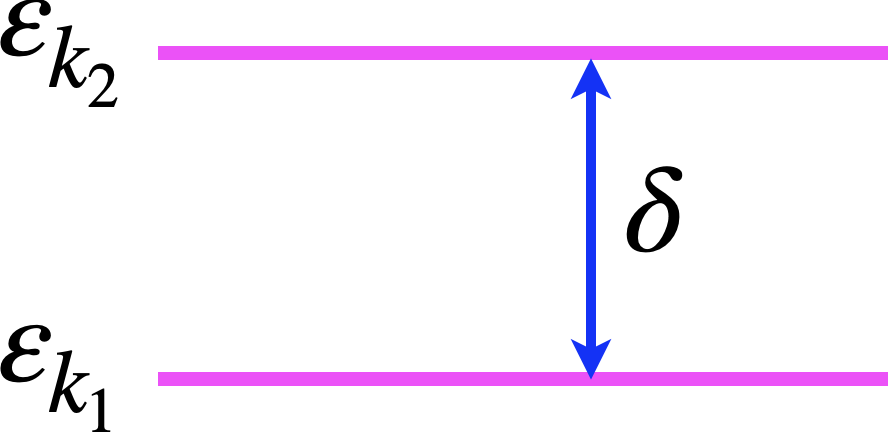}
\caption{\label{Second toy model} The two states $k_{1}$ and $k_{2}$ with the energy difference $\delta$ in the second toy model.} 
\end{figure} 

\begin{figure}
\includegraphics[scale=0.425]{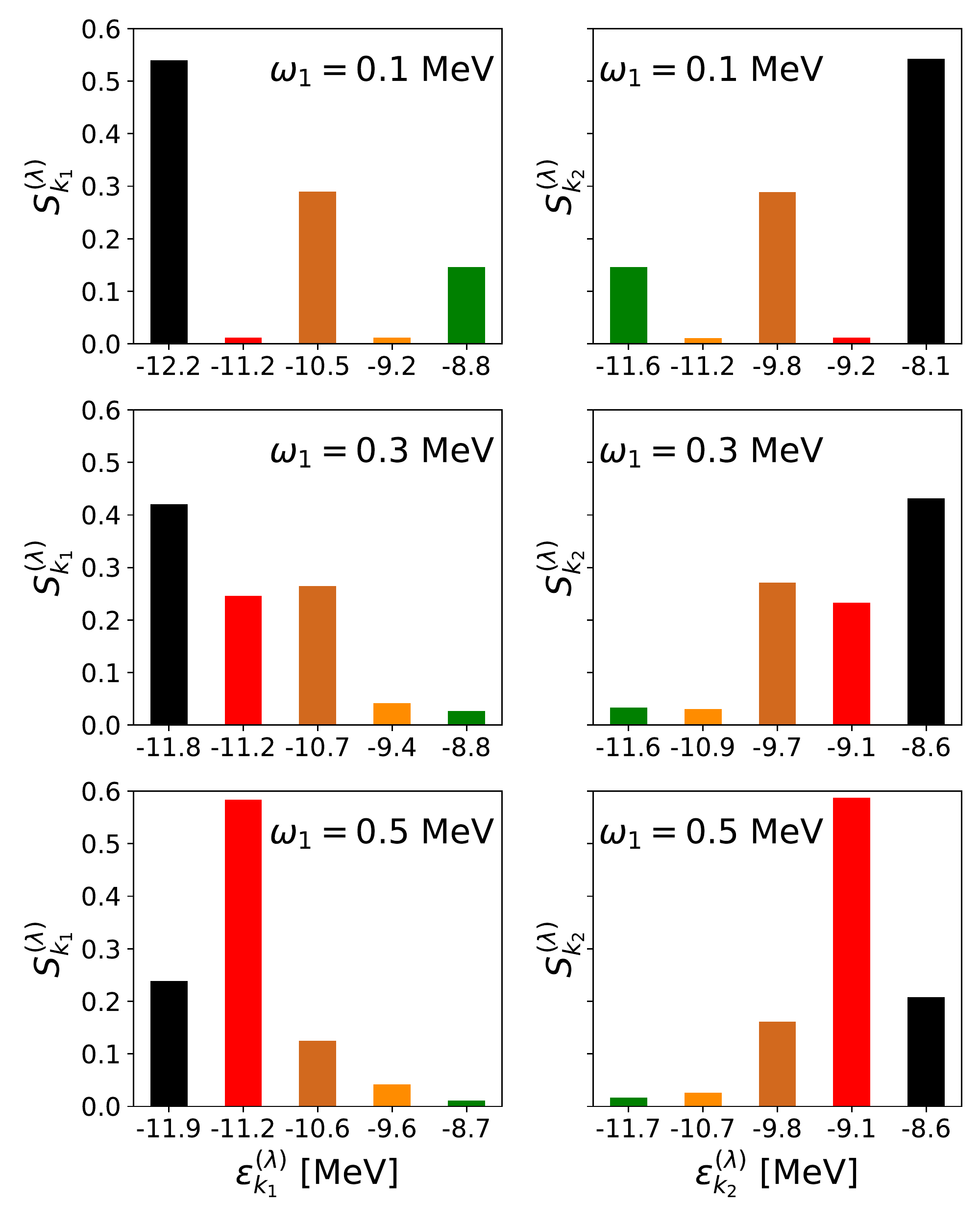}
\caption{\label{5-fragment}The evolution of the spectroscopic factors $S^{(\lambda)}_{k_{1}}$ (left panel) and $S^{(\lambda)}_{k_{2}}$ (right panel) with the phonon frequency $\omega_{1}$ at $T=1$ MeV. The parameter $\delta=2.0$ MeV.
} 
\end{figure}

\begin{figure}
\includegraphics[scale=0.425]{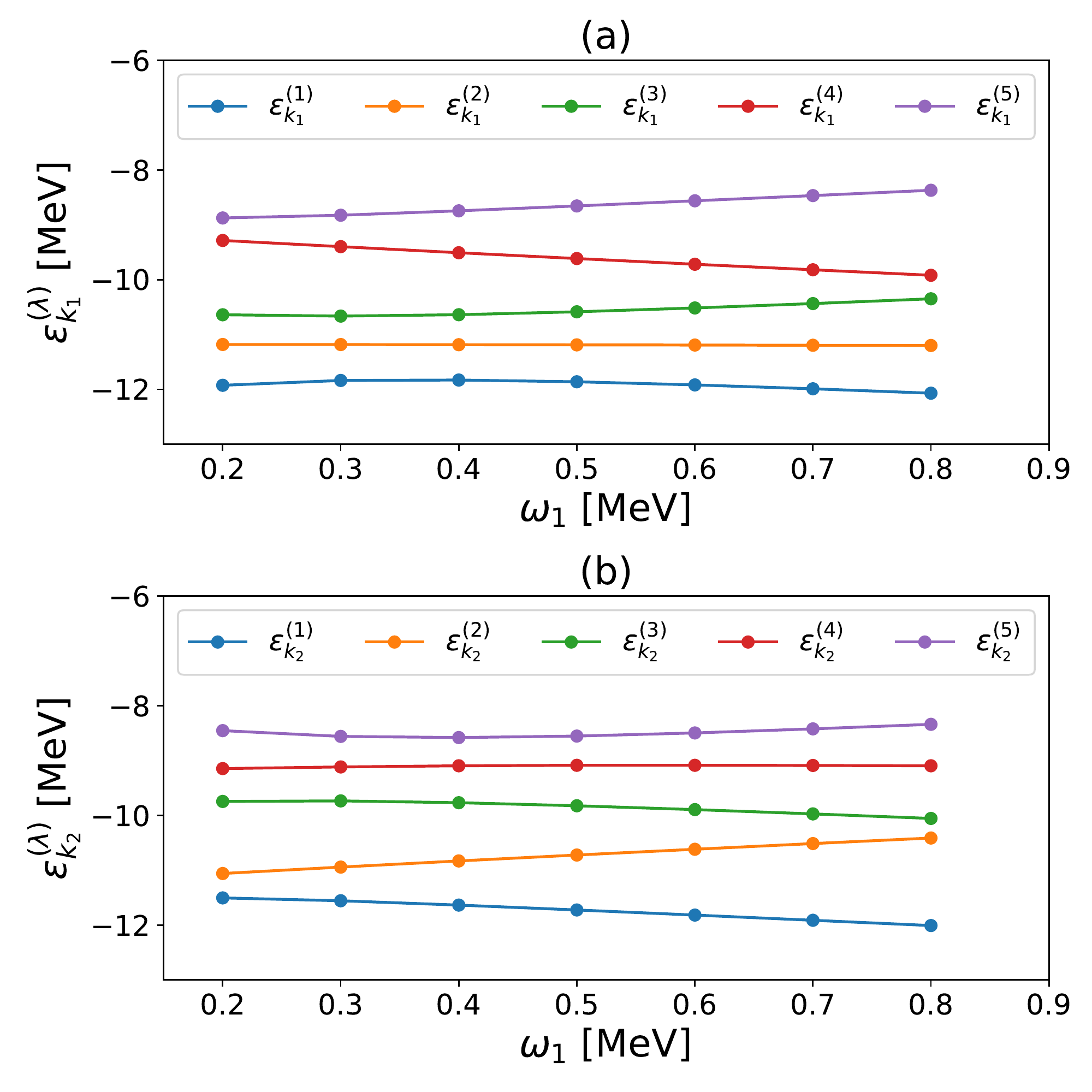}
\caption{\label{DeltaE of 5-fragment} The evolution of the energy fragments (a) $\varepsilon_{k_{1}}^{(\lambda)}$ and (b) $\varepsilon_{k_{2}}^{(\lambda)}$ with the phonon frequency $\omega_{1}$ at $T=1$ MeV. The parameter $\delta=2.0$ MeV.} 
\end{figure}

\begin{figure}
\includegraphics[scale=0.425]{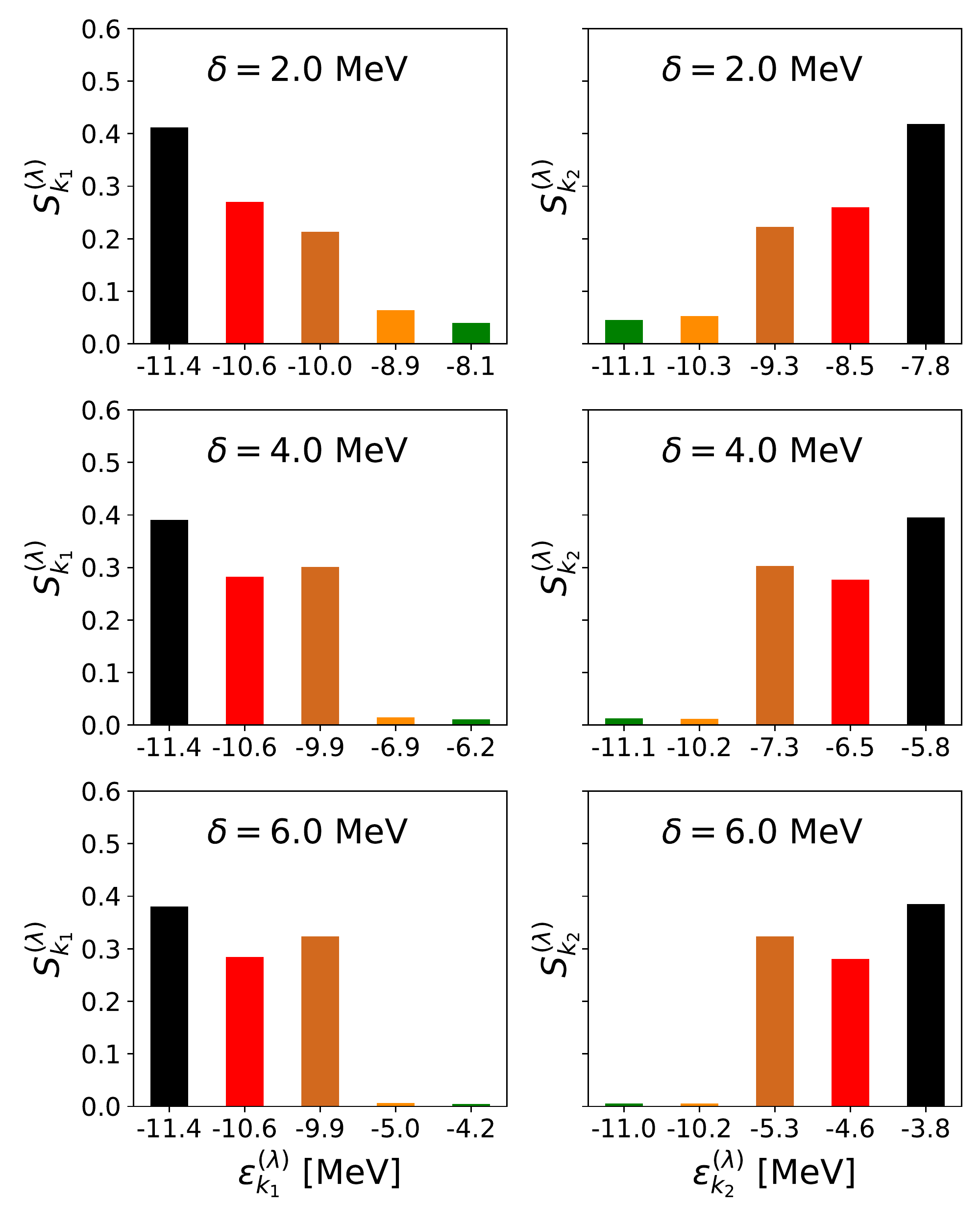}
\caption{\label{5-fragment-T_2MeV} The evolution of the energy fragments $\varepsilon_{k_{1}}^{(\lambda)}$ and $\varepsilon_{k_{2}}^{(\lambda)}$ with the parameter $\delta$ at $T=2$ MeV and the fixed phonon frequency $\omega_{1}=0.4$ MeV.} 
\end{figure}

\begin{figure}[h]
\includegraphics[scale=0.425]{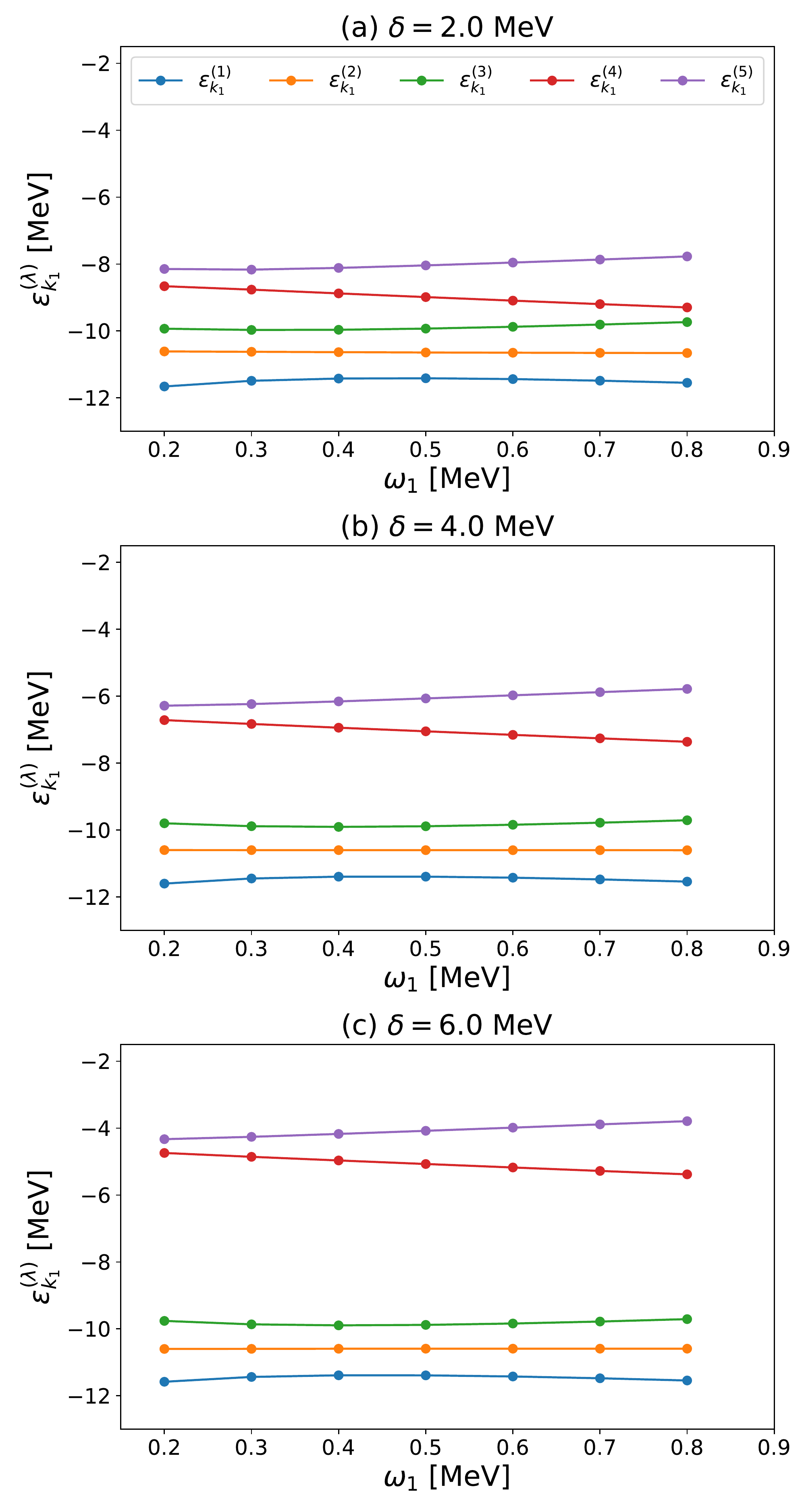}
\caption{\label{DeltaE of 5-fragment T 2 MeV} The evolution of the energy fragments $\varepsilon_{k_{1}}^{(\lambda)}$ with phonon frequency $\omega_{1}$ at $T=2$ MeV. Here the parameter $\delta$ takes the values: (a) 2.0 MeV, (b) 4.0 MeV, and (c) 6.0 MeV.} 
\end{figure}

In the second toy model, we add another state $k_{2}$ to the previous state $k_{1}$ with the energy difference $\delta\equiv \varepsilon_{k_{2}}-\varepsilon_{k_{1}}>0$, see Fig. \ref{Second toy model}. For the case of two-level system with one phonon mode, the two diagonal mass operators $\Sigma^{e}_{k_{1}}(\varepsilon)$ and $\Sigma^{e}_{k_{2}}(\varepsilon)$, respectively, take the form:
\begin{eqnarray}
\label{Sigma-e k1}\Sigma^{e}_{k_{1}}(\varepsilon)&=&g^{1}_{k_{1}k_{1}}g^{1\ast}_{k_{1}k_{1}}\frac{N(\omega_{1},T)+1-n(\varepsilon_{k_{1}}-\mu,T)}{\varepsilon-\varepsilon_{k_{1}}+\mu-\omega_{1}}\nonumber\\
&+&g^{1\ast}_{k_{1}k_{1}}g^{1}_{k_{1}k_{1}}\frac{n(\varepsilon_{k_{1}}-\mu,T)+N(\omega_{1},T)}{\varepsilon-\varepsilon_{k_{1}}+\mu+\omega_{1}}\nonumber\\
&+&g^{1}_{k_{1}k_{2}}g^{1\ast}_{k_{1}k_{2}}\frac{N(\omega_{1},T)+1-n(\varepsilon_{k_{2}}-\mu,T)}{\varepsilon-\varepsilon_{k_{2}}+\mu-\omega_{1}}\nonumber\\
&+&g^{1\ast}_{k_{2}k_{1}}g^{1}_{k_{2}k_{1}}\frac{n(\varepsilon_{k_{2}}-\mu,T)+N(\omega_{1},T)}{\varepsilon-\varepsilon_{k_{2}}+\mu+\omega_{1}}
\end{eqnarray}
and 
\begin{eqnarray}
\label{Sigma-e k2}\Sigma^{e}_{k_{2}}(\varepsilon)&=&g^{1}_{k_{2}k_{1}}g^{1\ast}_{k_{2}k_{1}}\frac{N(\omega_{1},T)+1-n(\varepsilon_{k_{1}}-\mu,T)}{\varepsilon-\varepsilon_{k_{1}}+\mu-\omega_{1}}\nonumber\\
&+&g^{1\ast}_{k_{1}k_{2}}g^{1}_{k_{1}k_{2}}\frac{n(\varepsilon_{k_{1}}-\mu,T)+N(\omega_{1},T)}{\varepsilon-\varepsilon_{k_{1}}+\mu+\omega_{1}}\nonumber\\
&+&g^{1}_{k_{2}k_{2}}g^{1\ast}_{k_{2}k_{2}}\frac{N(\omega_{1},T)+1-n(\varepsilon_{k_{2}}-\mu,T)}{\varepsilon-\varepsilon_{k_{2}}+\mu-\omega_{1}}\nonumber\\
&+&g^{1\ast}_{k_{2}k_{2}}g^{1}_{k_{2}k_{2}}\frac{n(\varepsilon_{k_{2}}-\mu,T)+N(\omega_{1},T)}{\varepsilon-\varepsilon_{k_{2}}+\mu+\omega_{1}}.
\end{eqnarray}
The solutions $\varepsilon_{k_{1}}^{(\lambda)}$ and $\varepsilon_{k_{2}}^{(\lambda)}$, where $\lambda = 1,\;2,\;...,\;5$, are obtained by diagonalizing the following two matrices:
\begin{equation}
\left( \begin{array}{ccccc}
\varepsilon_{k_{1}}-\mu & \xi^{1(+1)}_{k_{1}k_{1}} & \cdots & \cdots & \xi^{1(-1)}_{k_{2}k_{1}} \\ 
\xi^{1(+1)\ast}_{k_{1}k_{1}} & \varepsilon_{k_{1}}-\mu-\omega_{1} & 0 & \cdots & 0 \\ 
\vdots & 0 & \ddots & \cdots & 0 \\ 
\vdots & \vdots & \vdots & \ddots & 0 \\ 
\xi^{1(-1)\ast}_{k_{2}k_{1}} & 0 & 0 & 0 & \varepsilon_{k_{2}}-\mu+\omega_{1}
\end{array}  \right)
\end{equation}
and
\begin{equation}
\left( \begin{array}{ccccc}
\varepsilon_{k_{2}}-\mu & \xi^{1(+1)}_{k_{1}k_{2}} & \cdots & \cdots & \xi^{1(-1)}_{k_{2}k_{2}} \\ 
\xi^{1(+1)\ast}_{k_{1}k_{2}} & \varepsilon_{k_{1}}-\mu-\omega_{1} & 0 & \cdots & 0 \\ 
\vdots & 0 & \ddots & \cdots & 0 \\ 
\vdots & \vdots & \vdots & \ddots & 0 \\ 
\xi^{1(-1)\ast}_{k_{2}k_{2}} & 0 & 0 & 0 & \varepsilon_{k_{2}}-\mu+\omega_{1}
\end{array}  \right).
\end{equation}
According to Eq. \eqref{Spectroscopic factor}, the corresponding spectroscopic factors $S_{k_{1}}^{(\lambda)}$ and $S_{k_{2}}^{(\lambda)}$ are determined via
\begin{eqnarray}
S_{k_{1}}^{(\lambda)}=\left[1-\frac{d}{d\varepsilon}\Sigma^{e}_{k_{1}}(\varepsilon)\right]^{-1}_{\varepsilon=\varepsilon_{k_{1}}^{(\lambda)}}
\end{eqnarray}
and 
\begin{eqnarray}
S_{k_{2}}^{(\lambda)}=\left[1-\frac{d}{d\varepsilon}\Sigma^{e}_{k_{2}}(\varepsilon)\right]^{-1}_{\varepsilon=\varepsilon_{k_{2}}^{(\lambda)}},
\end{eqnarray}
respectively. As before, we suppose the state $k_{1}$ to be the state $1\text{f}_{7/2}$ in the $^{70}\text{Ni}$ nucleus and set all the phonon vertices equal to 0.2. We first computed the spectroscopic factors $S_{k_{1}}^{(\lambda)}$ and $S_{k_{2}}^{(\lambda)}$ for the fixed temperature $T=1$ MeV, at which the chemical potential $\mu=-6.114$ MeV and the RMF energy $\varepsilon_{k_{1}}=-17.296$ MeV. The energy of the second state $k_{2}$ is also fixed according to $\varepsilon_{k_{2}}=\varepsilon_{k_{1}}+\delta$, where $\delta=2.0$ MeV. A smaller (larger) value of $\delta$ indicates a smaller (larger) energy distance of the state $k_{2}$ with respect to the state $k_{1}$. The left (right) panel of Fig. \ref{5-fragment} displays the evolution of the spectroscopic factor $S^{(\lambda)}_{k_{1}}$ ($S^{(\lambda)}_{k_{2}}$) with the phonon frequency $\omega_{1}$ at $T=1$ for the case of $\delta=2.0$ MeV. From Fig. \ref{5-fragment}, one obtains the critical phonon frequency $\omega_{1}^{\text{crit}}$ of 0.3 MeV. This is in accordance with the energy fragments $\varepsilon_{k_{1}}^{(1,2,3)}$ ($\varepsilon_{k_{2}}^{(3,4,5)}$) being close to each other at $\omega_{1}=0.3$ MeV, as demonstrated in Fig. \ref{DeltaE of 5-fragment}a (\ref{DeltaE of 5-fragment}b). A comparison between the two panels of Fig. \ref{5-fragment} shows an apparent mirror symmetry between both spectroscopic factor distributions. This symmetry occurs as a consequence of the single-valued phonon vertices, leading to the symmetry between the mass operators $\Sigma^{e}_{k_{1}}$ and $\Sigma^{e}_{k_{2}}$ with respect to the interchange $k_{1}\leftrightarrow k_{2}$ (see Eqs. \eqref{Sigma-e k1} and \eqref{Sigma-e k2}). Next, we increased the temperature $T$ to 2 MeV and  let the parameter $\delta$ take the values of 2.0 MeV, 4.0 MeV, and 6.0 MeV. Fig. \ref{5-fragment-T_2MeV} displays the distribution of the spectroscopic factors $S_{k_{1}}^{(\lambda)}$ and $S_{k_{2}}^{(\lambda)}$ at $T=2$ MeV for the various values of $\delta$ and the fixed phonon frequency $\omega_{1}=0.4$ MeV. Fig. \ref{5-fragment-T_2MeV} demonstrates that the  mirror symmetry between the two spectroscopic strength distributions persists for larger $\delta$ and higher temperature $T$. For several values of the parameter $\delta$, the evolution of the energy fragments $\varepsilon_{k_{1}}^{(\lambda)}$ with the phonon frequency $\omega_{1}$ at $T=2$ MeV is shown by Fig. \ref{DeltaE of 5-fragment T 2 MeV}. As one can see from Fig. \ref{DeltaE of 5-fragment T 2 MeV}, the critical phonon frequency $\omega_{1}^{\text{crit}}$ remains constant, i.e., 0.4 MeV, as one varies the parameter $\delta$. Analogously to the first toy model, the critical phonon frequency $\omega_{1}^{\text{crit}}$ increases roughly by 0.1 MeV per 1 MeV temperature increase, regardless the value of $\delta$. From Fig. \ref{DeltaE of 5-fragment T 2 MeV}, one observes that the energy fragments $\varepsilon_{k_{1}}^{(1,2,3)}$ becoming more and more separated as the value of parameter $\delta$ increases. As a result, the sum of the spectroscopic factors $S^{(1,2,3)}_{k_{1}}$ gradually exhausts the sum rule \eqref{Sum Rule}, as it follows from Fig. \ref{5-fragment-T_2MeV}. 

\begin{figure}
\includegraphics[scale=0.20]{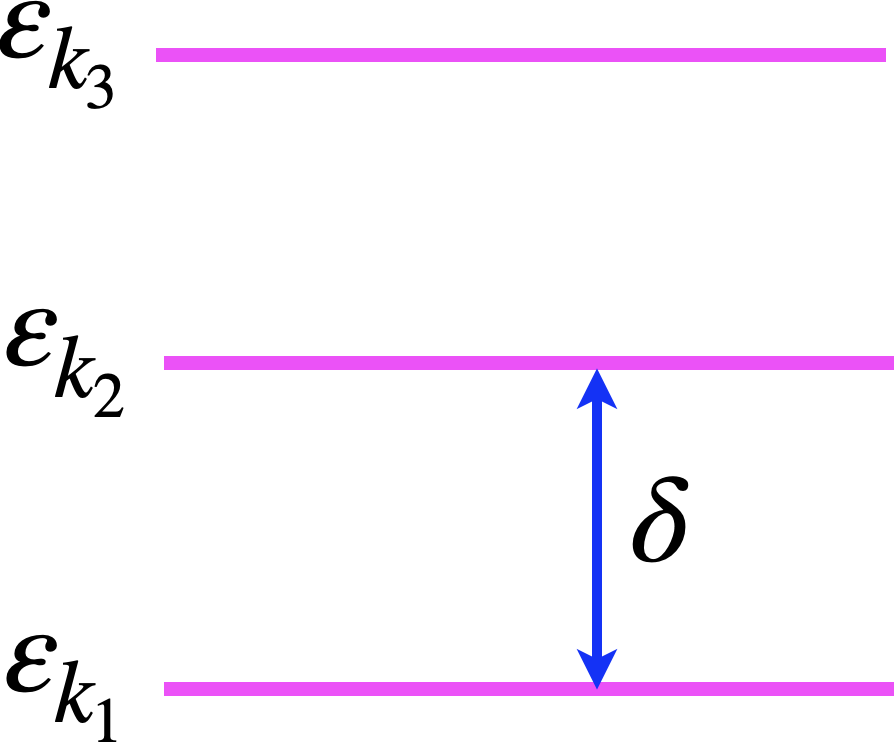}
\caption{\label{Third toy model} The three states $k_{1}$, $k_{2}$, and $k_{3}$ in the third toy model.The parameter $\delta$ refers to the energy difference between the states $k_{1}$ and $k_{2}$.} 
\end{figure} 

\begin{figure}
\includegraphics[scale=0.5]{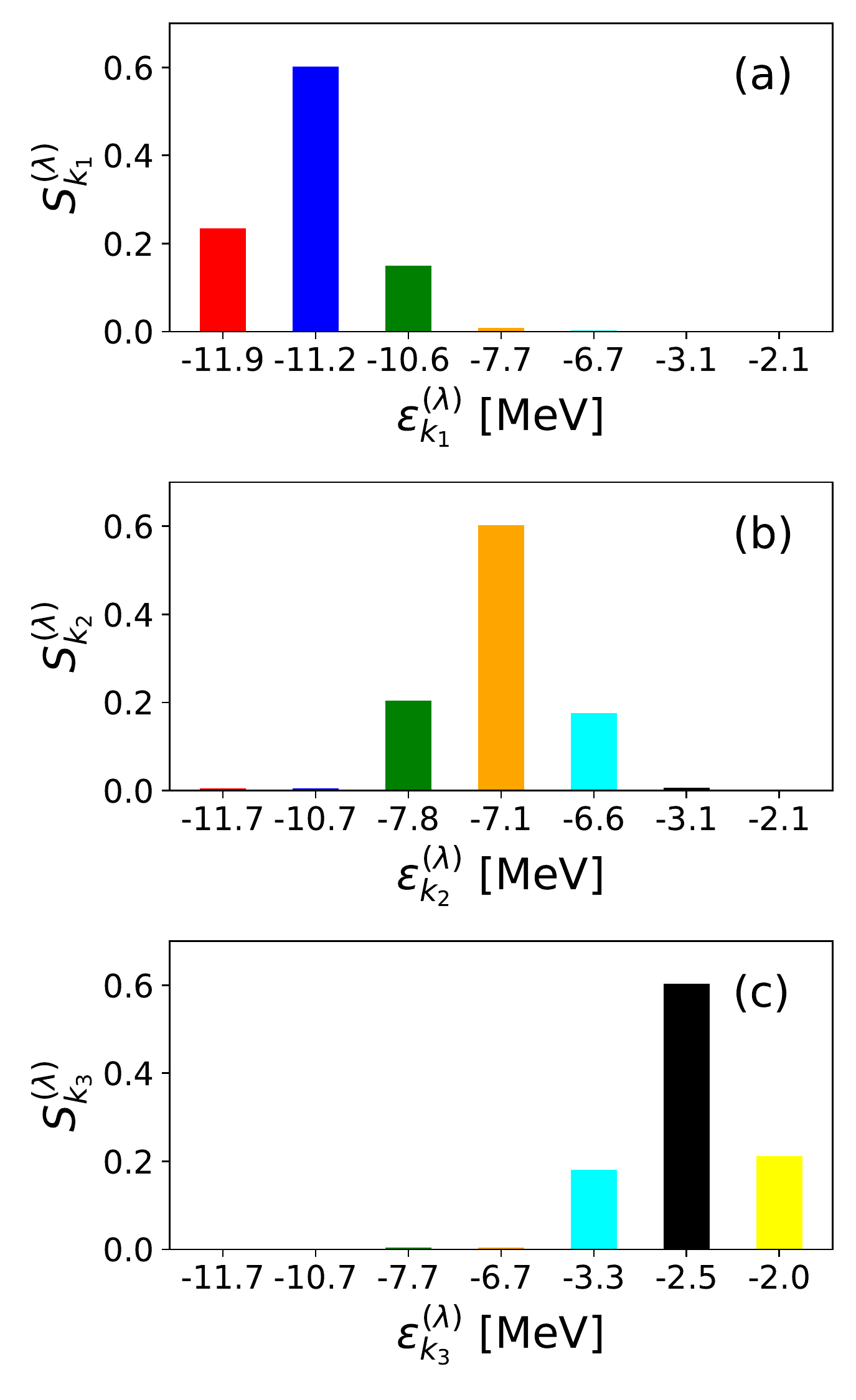}
\caption{\label{7-fragment_delta4} The spectroscopic factor distributions of the energy fragments (a) $\varepsilon_{k_{1}}^{(\lambda)}$, (b) $\varepsilon_{k_{2}}^{(\lambda)}$, and (c) $\varepsilon_{k_{3}}^{(\lambda)}$, where $\lambda = 1,\;...,\;7$. Here, the parameter $\delta=4.0$ and the phonon frequency $\omega_{1}=0.5$ MeV.} 
\end{figure} 

\begin{figure}
\includegraphics[scale=0.5]{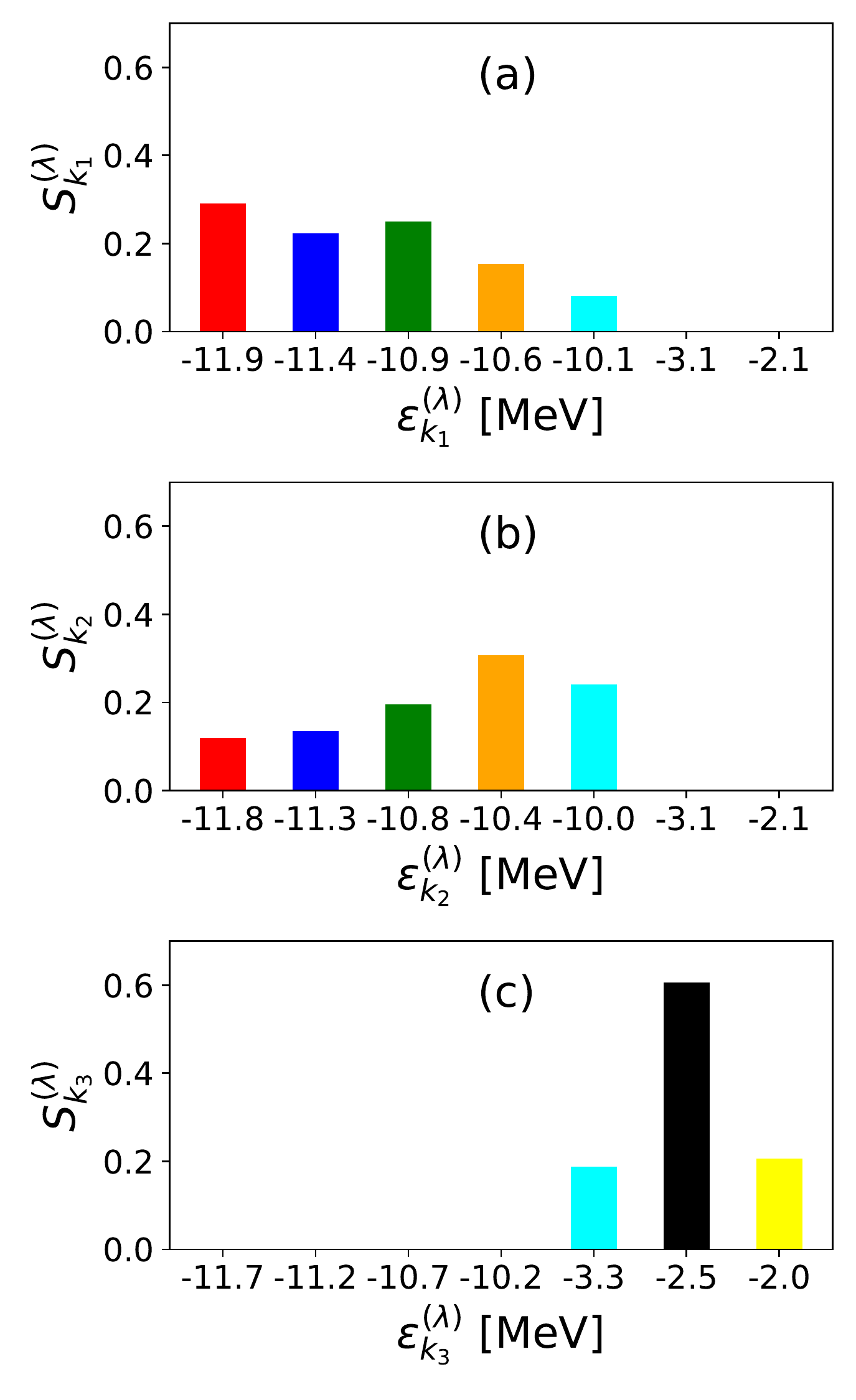}
\caption{\label{7-fragment_delta05} Same as Fig. \ref{7-fragment_delta4}, but for  $\delta=0.5$ MeV.} 
\end{figure}

\begin{figure}
\includegraphics[scale=0.5]{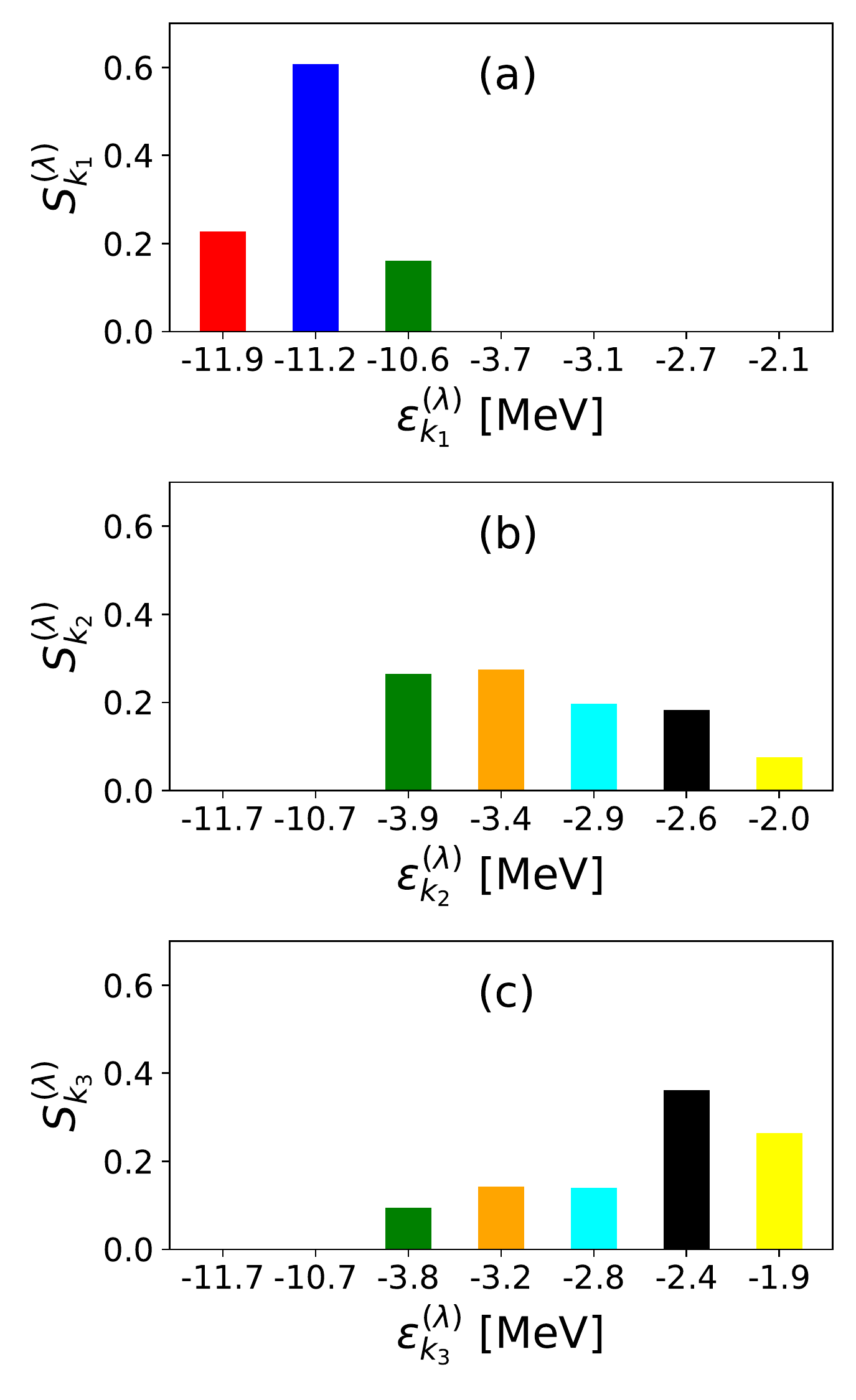}
\caption{\label{7-fragment_delta8} Same as Fig. \ref{7-fragment_delta4}, but for  $\delta=8.0$ MeV.} 
\end{figure}

In the third toy model, we consider a system which consists of three states $k_{1}$, $k_{2}$, and $k_{3}$,  as shown in Fig. \ref{Third toy model}, and one phonon mode with the frequency $\omega_{1}$. The energies of both states $k_{1}$ and $k_{3}$ are fixed, while the state $k_{2}$ is separated by the energy $\delta$ from the state $k_{1}$. As before, we associate the states $k_{1}$ and $k_{3}$ with the states $1\text{f}_{7/2}$ and $2\text{p}_{1/2}$ in the $^{70}\text{Ni}$ nucleus. At $T=1$ MeV, the RMF energy of the state $2\text{p}_{1/2}$ is $-8.717$ MeV. We again set all the phonon vertices equal to 0.2. Since the energy difference between the states $k_{1}$ and $k_{3}$ is around 8.0 MeV, it is instructive to consider several values of $\delta$, which we set equal to 0.5, 4.0, and 8.0 MeV. 
In the case of $\delta=4.0$ MeV, when the state $k_{2}$ is located approximately in the middle between the states $k_{1}$ and $k_{3}$, all of the mean-field states exhibit a similar fragmentation pattern in all temperature regimes and at various phonon frequencies. For example, all of the mean-field states are weakly fragmented at $T=1$ MeV and with the phonon frequency $\omega_{1}=0.5$ MeV, as shown in Fig. \ref{7-fragment_delta4}. In contrast, for the case of $\delta=0.5$ MeV, the states $k_{1}$ and $k_{2}$ are strongly fragmented, whereas the state $k_{3}$ is weakly fragmented, as displayed in Fig. \ref{7-fragment_delta05}. Analogously to the case of $\delta=0.5$ MeV, for $\delta=8.0$ MeV the states $k_{2}$ and $k_{3}$ are strongly fragmented, leaving the state $k_{1}$ weakly fragmented, as shown in Fig. \ref{7-fragment_delta8}. From the second and third toy models, one concludes that the degree of fragmentation of each state is sensitive to the relative distance between the neighboring mean-field states, if they are connected by the PVC mechanism. As it is shown in Figs. \ref{5-fragment}, \ref{5-fragment-T_2MeV}, and \ref{7-fragment_delta4}, the neighboring mean-field states with the same relative distances to their PVC partners exhibit the same fragmentation pattern. Overall, the neighboring states, which are closer to each other, are stronger fragmented, as demonstrated in Figs. \ref{7-fragment_delta05} and \ref{7-fragment_delta8}.

\begin{table*}
\caption{The temperature evolution of the dominant fragments of the continuum $3\text{p}_{1/2}$ and $3\text{p}_{3/2}$ states in $^{70,74}\text{Ni}$ isotopes. Here $\varepsilon_{k}^{\text{dom}}$ and $\text{S}_{k}^{\text{dom}}$ represent the energy and the corresponding spectroscopic factor for each dominant fragment. }
\begin{ruledtabular}
\begin{tabular}{ccccccccccc}
 & \multicolumn{2}{c}{$T=0$} &  \multicolumn{2}{c}{$T=1$ MeV} & \multicolumn{2}{c}{$T=2$ MeV} & \multicolumn{2}{c}{$T=3$ MeV} & \multicolumn{2}{c}{$T=4$ MeV} \\ 
Orbital & $\varepsilon_{k}^{\text{dom}}$ & $S_{k}^{\text{dom}}$ & $\varepsilon_{k}^{\text{dom}}$ & $S_{k}^{\text{dom}}$ & $\varepsilon_{k}^{\text{dom}}$ & $S_{k}^{\text{dom}}$ & $\varepsilon_{k}^{\text{dom}}$ & $S_{k}^{\text{dom}}$ & $\varepsilon_{k}^{\text{dom}}$ & $S_{k}^{\text{dom}}$ \\ 
 & [MeV] & & [MeV] & & [MeV] & & [MeV] & & [MeV] &  \\ 
\hline 
\multicolumn{11}{l}{$^{70}\text{Ni}$}  \\ 
$3\text{p}_{1/2}$ & 3.226 & 0.918 & 2.889 & 0.303 & 2.836 & 0.317 & 2.681 & 0.342 & 2.583 & 0.643 \\ 
 &  &  & 3.204 & 0.673 & 3.162 & 0.514 & 3.055 & 0.505 & 3.021 & 0.285 \\ 
$3\text{p}_{3/2}$ & 3.119 & 0.954 & 2.766 & 0.534 & 2.795 & 0.565 & 2.677 & 0.587 & 2.491 & 0.866 \\ 
 &  &  & 3.054 & 0.333 & 3.110 & 0.286 &  &  &  &  \\ 
\hline 
\multicolumn{11}{l}{$^{74}\text{Ni}$}  \\ 
$3\text{p}_{1/2}$ & 3.082 & 0.348 & 2.992 & 0.885 & 2.948 & 0.245 & 2.750 & 0.386 & 2.213 & 0.379 \\ 
 & 3.106 & 0.602 &  &  & 3.053 & 0.536 & 2.938 & 0.186 & 2.689 & 0.569 \\ 
$3\text{p}_{3/2}$ & 2.979 & 0.950 & 2.884 & 0.967 & 2.624 & 0.292 & 2.538 & 0.348 & 2.209 & 0.622 \\ 
 &  &  &  &  & 2.825 & 0.662 & 2.707 & 0.175 & 2.508 & 0.279 \\ 
\end{tabular} 
\label{Continuum states}
\end{ruledtabular}
\end{table*}


\begin{figure*}
\begin{subfigure}{0.3\textwidth}
\centering
\includegraphics[scale=0.60]{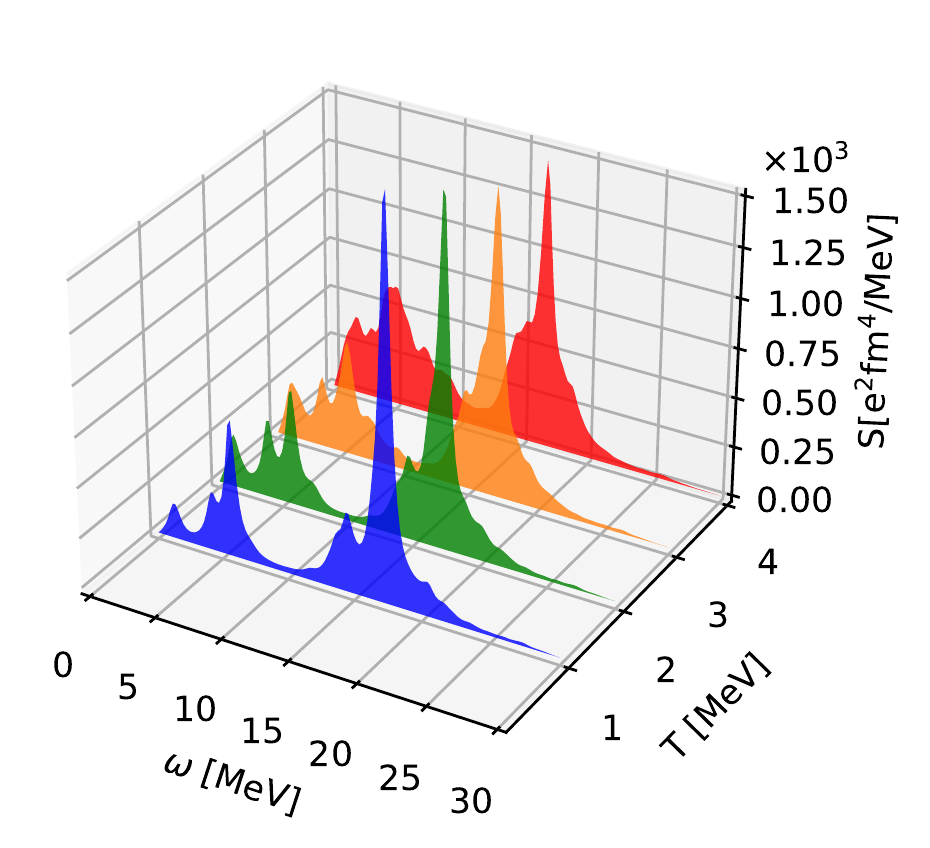} 
\caption{}
\label{Ni68_2+_Phonon} 
\end{subfigure}
\begin{subfigure}{0.3\textwidth}
\centering
\includegraphics[scale=0.60]{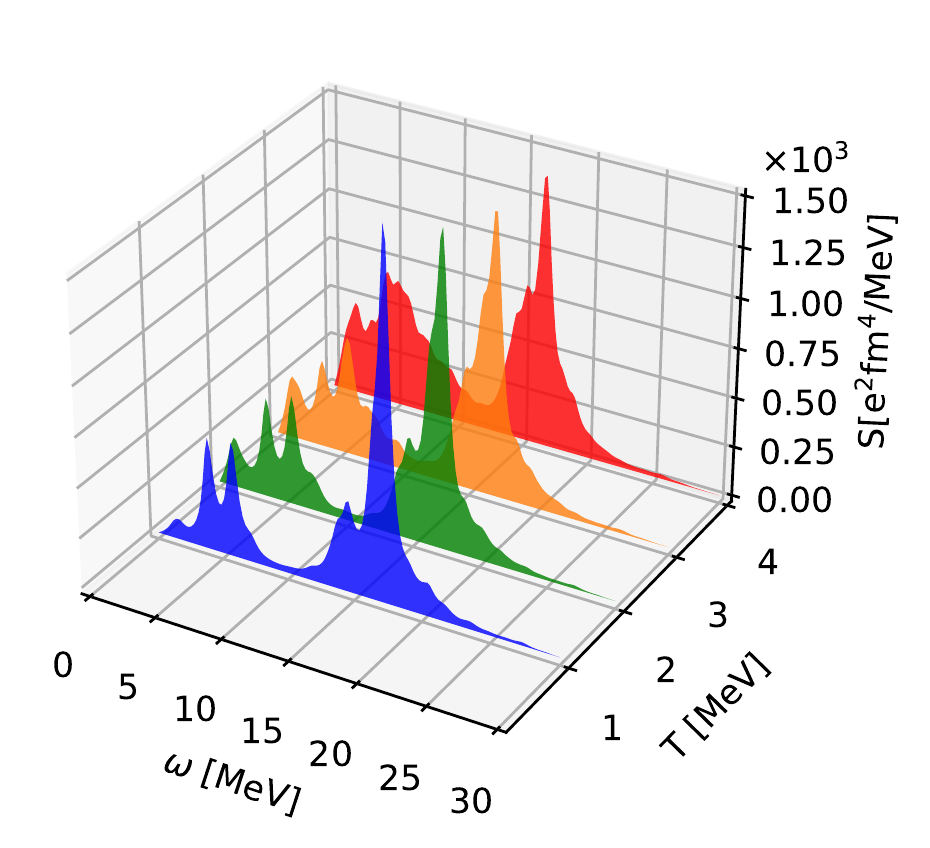}
\caption{}
\label{Ni70_2+_Phonon}
\end{subfigure}
\begin{subfigure}{0.3\textwidth}
\centering
\includegraphics[scale=0.60]{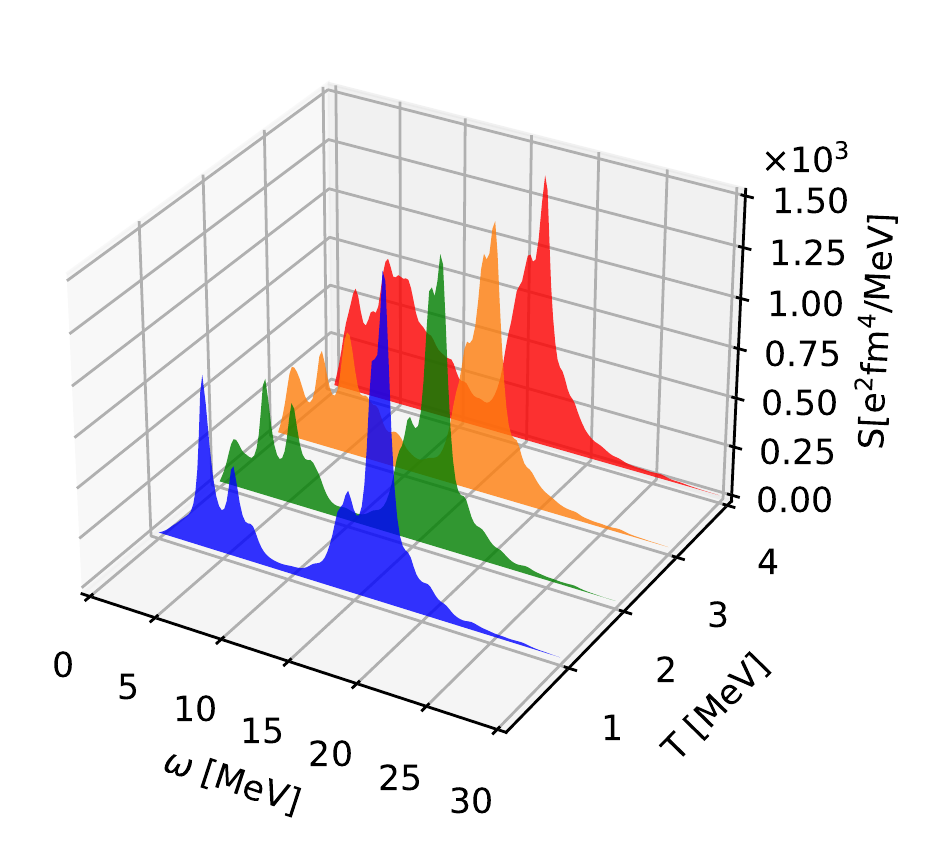}
\caption{}
\label{Ni72_2+_Phonon} 
\end{subfigure}
\begin{subfigure}{0.3\textwidth}
\centering
\includegraphics[scale=0.60]{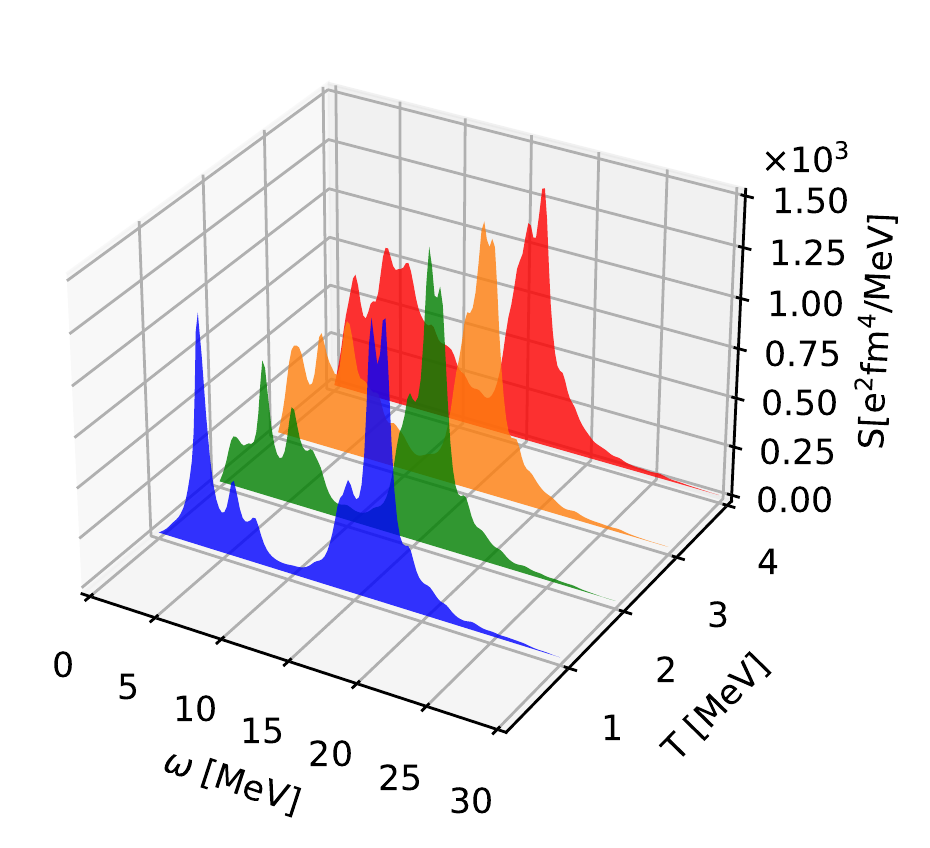}
\caption{} 
\label{Ni74_2+_Phonon} 
\end{subfigure}
\begin{subfigure}{0.3\textwidth}
\centering
\includegraphics[scale=0.60]{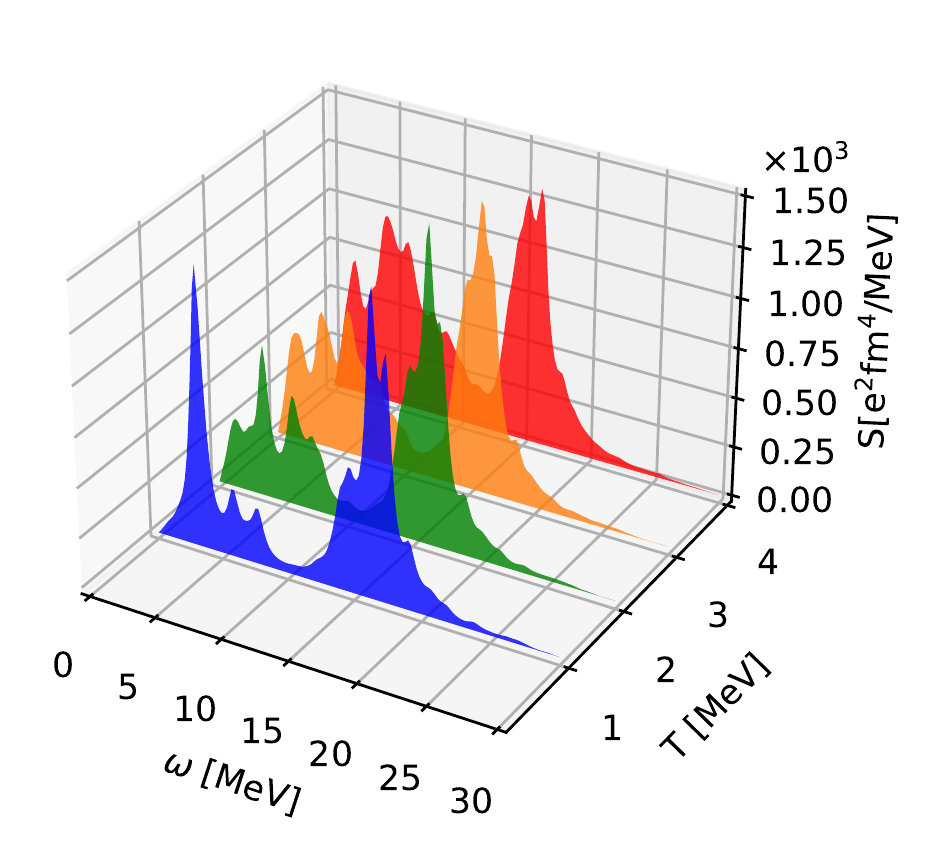}
\caption{} 
\label{Ni76_2+_Phonon} 
\end{subfigure}
\begin{subfigure}{0.3\textwidth}
\centering
\includegraphics[scale=0.60]{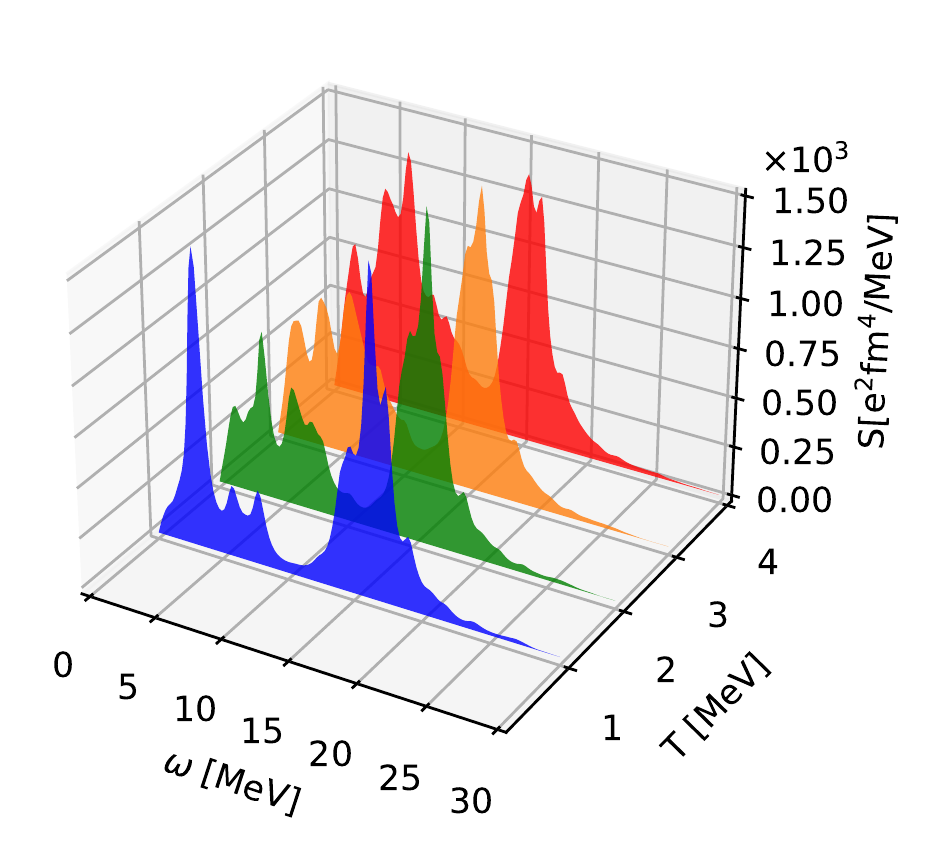}
\caption{} 
\label{Ni78_2+_Phonon} 
\end{subfigure}
\caption{Temperature dependence of the quadrupole ($2^{+}$) phonon strength distributions in (a) $^{68}\text{Ni}$, (b) $^{70}\text{Ni}$, (c) $^{72}\text{Ni}$, (d) $^{74}\text{Ni}$, (e) $^{76}\text{Ni}$, and (f) $^{78}\text{Ni}$.}
\end{figure*}


\begin{figure*}
\begin{subfigure}{0.3\textwidth}
\centering
\includegraphics[scale=0.60]{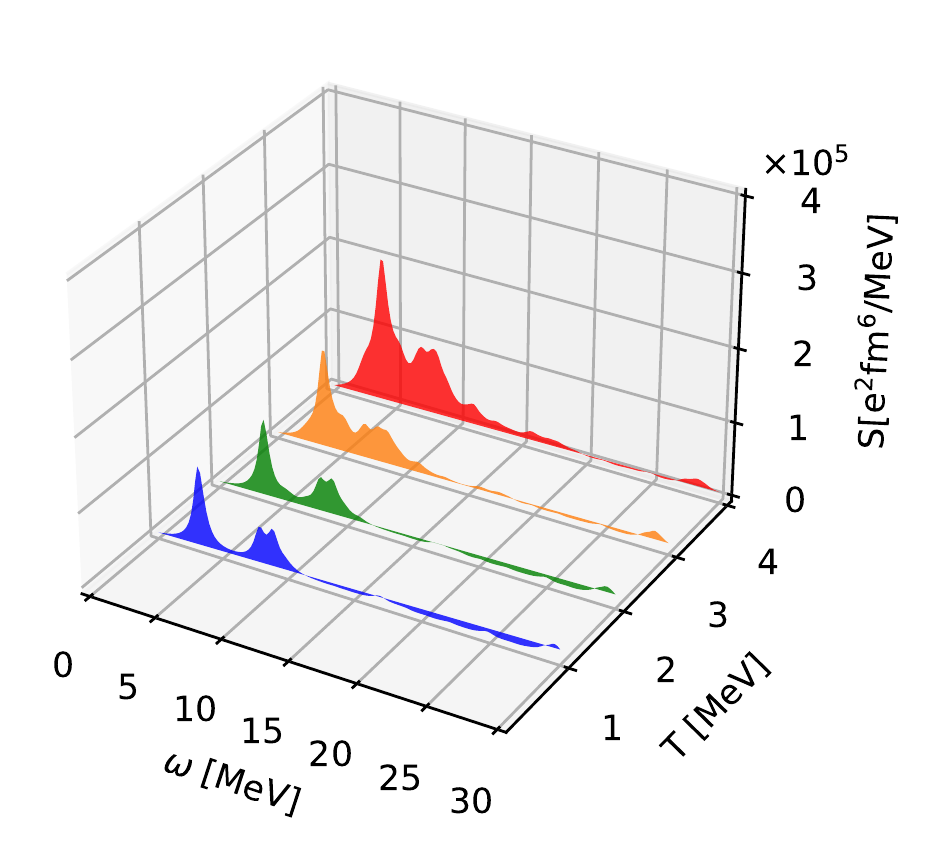} 
\caption{}
\label{Ni68_3-_Phonon} 
\end{subfigure}
\begin{subfigure}{0.3\textwidth}
\centering
\includegraphics[scale=0.60]{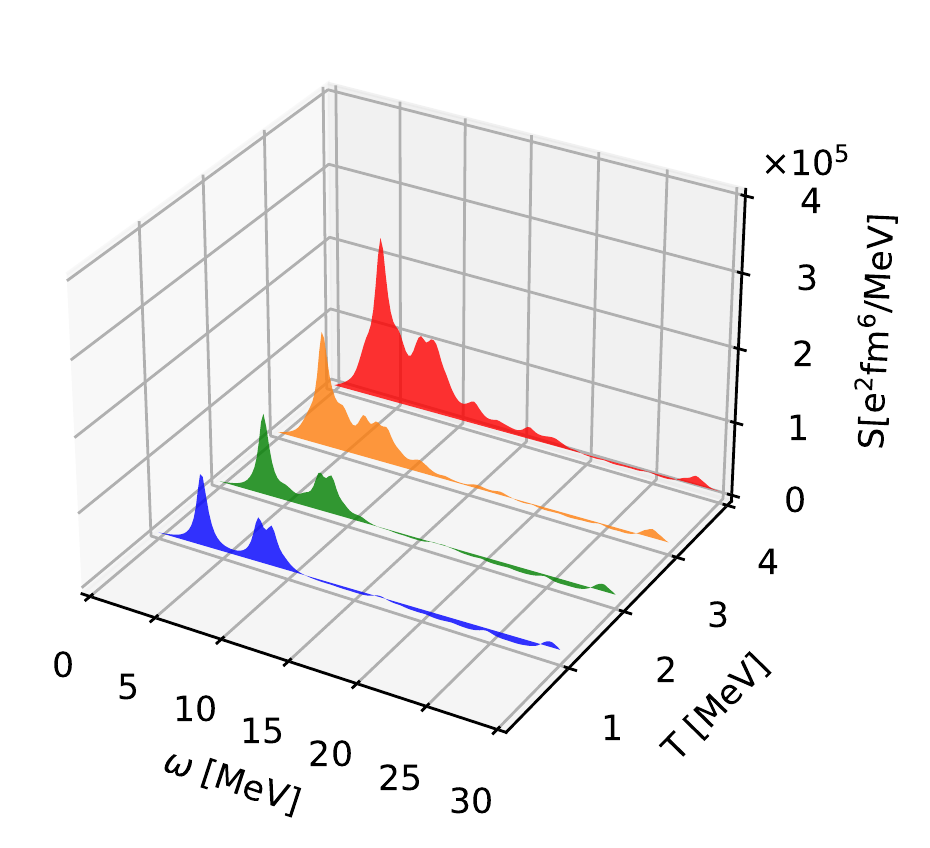}
\caption{}
\label{Ni70_3-_Phonon}
\end{subfigure}
\begin{subfigure}{0.3\textwidth}
\centering
\includegraphics[scale=0.60]{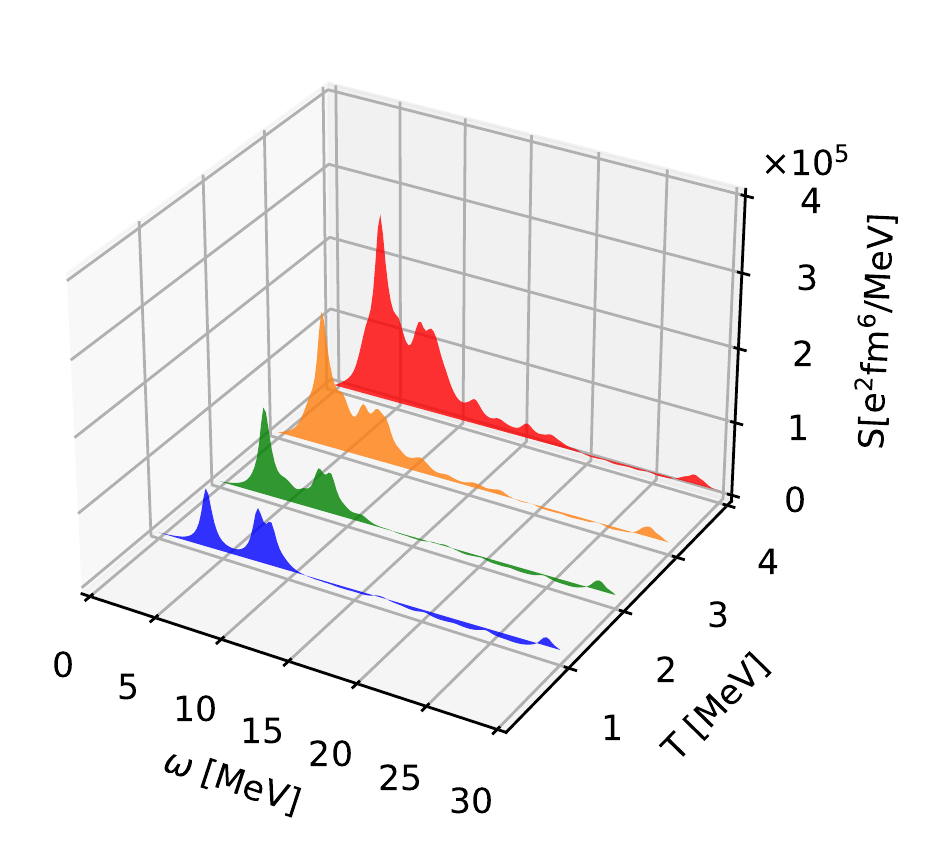}
\caption{}
\label{Ni72_3-_Phonon} 
\end{subfigure}
\begin{subfigure}{0.3\textwidth}
\centering
\includegraphics[scale=0.60]{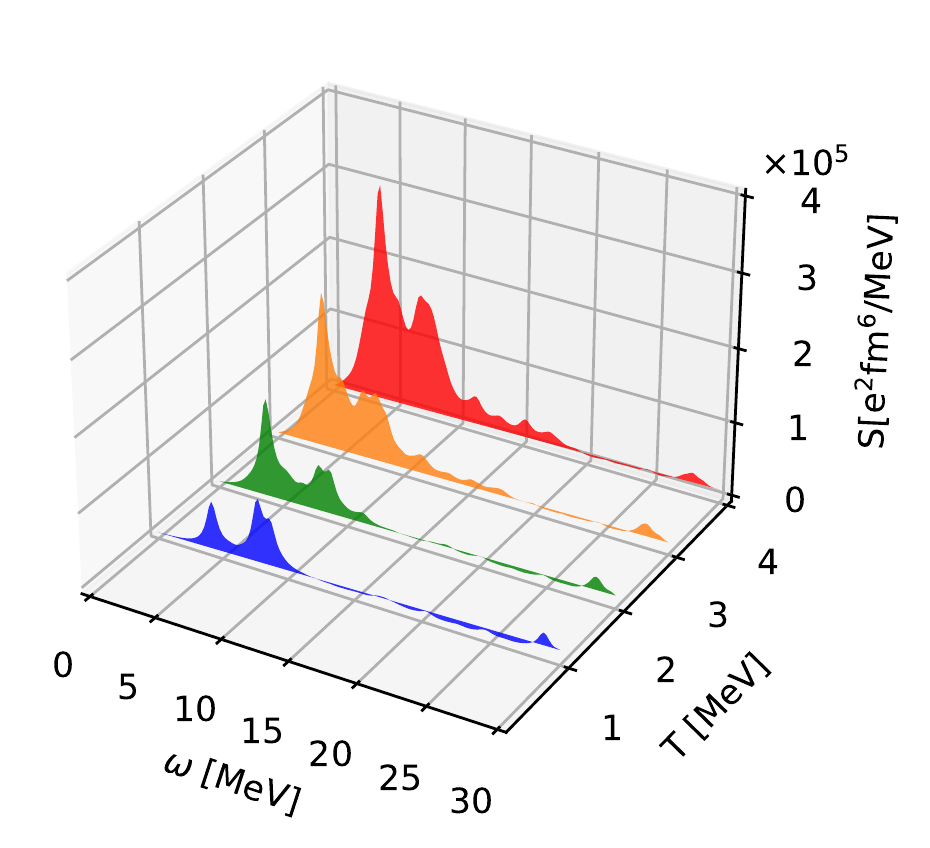}
\caption{} 
\label{Ni74_3-_Phonon} 
\end{subfigure}
\begin{subfigure}{0.3\textwidth}
\centering
\includegraphics[scale=0.60]{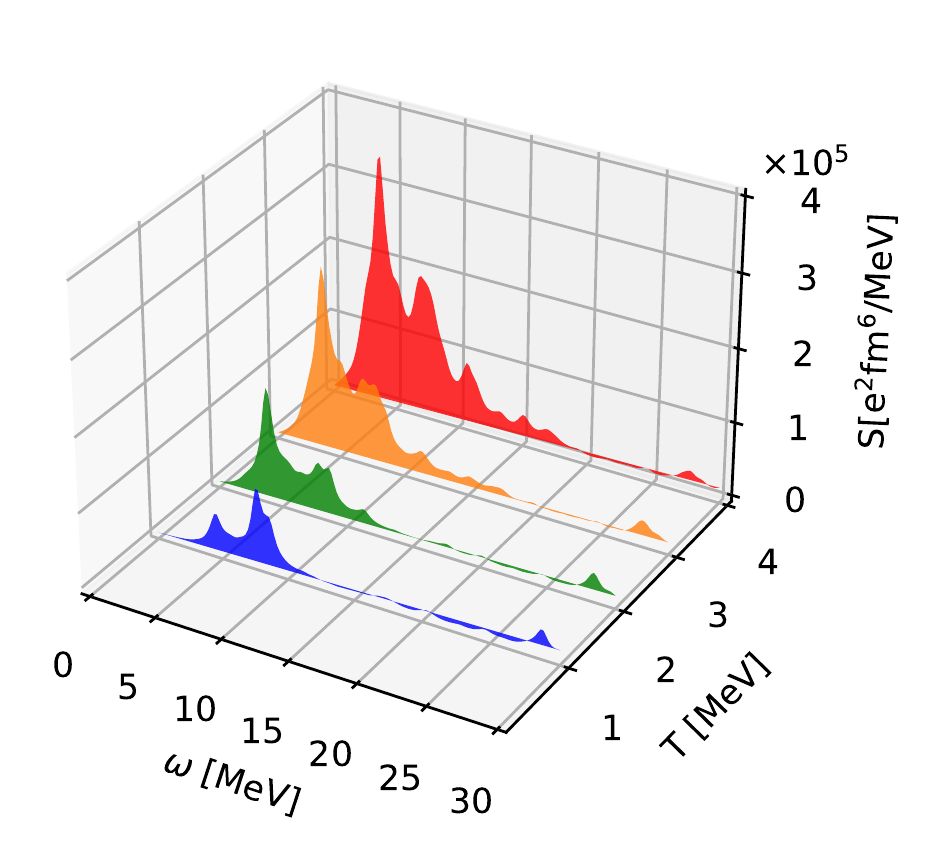}
\caption{} 
\label{Ni76_3-_Phonon} 
\end{subfigure}
\begin{subfigure}{0.3\textwidth}
\centering
\includegraphics[scale=0.60]{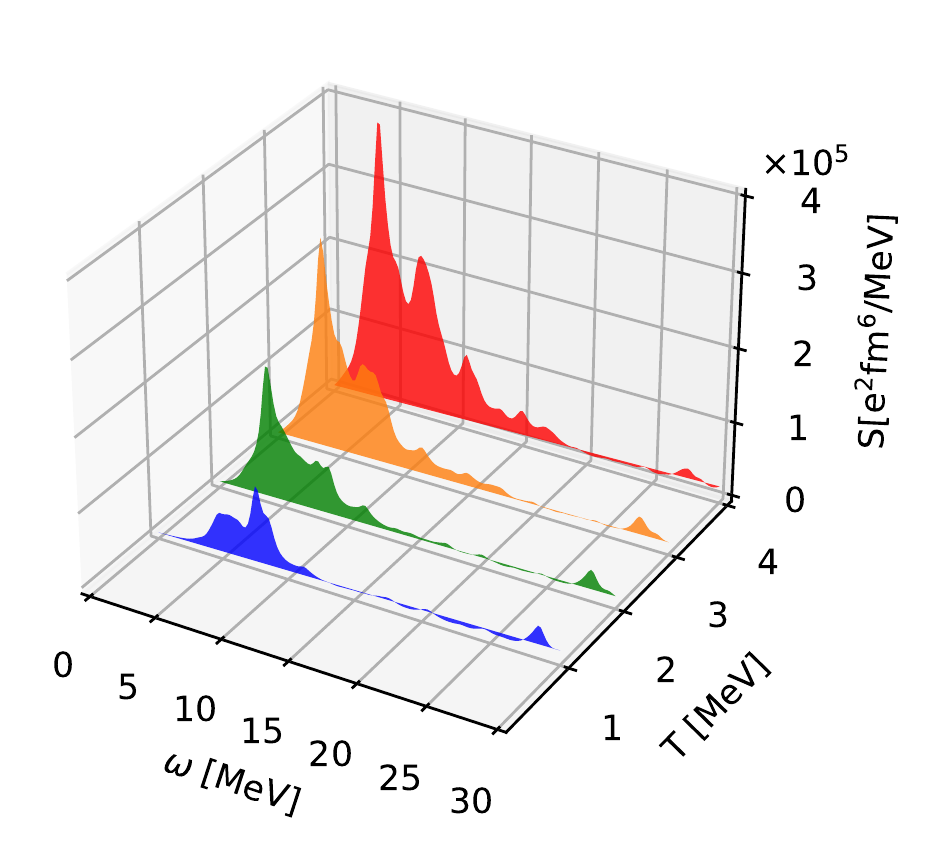}
\caption{} 
\label{Ni78_3-_Phonon} 
\end{subfigure}
\caption{Temperature dependence of the octupole ($3^{-}$) phonon strength distributions in (a) $^{68}\text{Ni}$, (b) $^{70}\text{Ni}$, (c) $^{72}\text{Ni}$, (d) $^{74}\text{Ni}$, (e) $^{76}\text{Ni}$, and (f) $^{78}\text{Ni}$.}
\end{figure*}

Summarizing our qualitative study within the simplistic models, we conclude:
\begin{enumerate}[(i)]
\item The number $N_{\lambda}$ of fragments generated by the PVC for each phonon mode coupled to a single-particle state $k$ satisfies the following equation:
\begin{equation}
N_{\lambda}=2N_{k_{3}}+1,
\end{equation}
where $N_{k_{3}}$ denotes the number of intermediate states $k_{3}$ in the mass operator $\Sigma^{e}$ (\ref{Sigma e}).
\item The energy differences between a specific state $k$ and its neighboring states determine the degree of fragmentation of the state $k$. The state $k$ is strongly fragmented if the energy differences are small.
\item The low-frequency phonons play the most important role in the fragmentation of the single-particle states.
\end{enumerate}

\subsection{The influence of phonons }

Our qualitative study within the toy models discussed above is very instructive for understanding the general trends of the PVC mechanism, however, the toy models can not explain the whole variety of fragmentation patterns. The deficiency of the simplistic toy models is that they do not take into account  the spin and parity of the mean-field states and of the various phonon modes into consideration. For example, the parity of the mean-field hole states $2\text{p}_{1/2}$, $2\text{p}_{3/2}$, $1\text{f}_{5/2}$, and $1\text{f}_{7/2}$ is negative. Consequently, the transitions between these mean-field hole states occur due to coupling to the positive-parity phonons, e.g., $2^{+}$ and $4^{+}$ phonons. Some mean-field particle states, such as $2\text{d}_{5/2}$, $3\text{s}_{1/2}$, $2\text{d}_{3/2}$, and $1\text{g}_{7/2}$, have a positive parity, whereas the states $3\text{p}_{3/2}$ and $3\text{p}_{1/2}$ lying in the continuum have a negative parity. In addition to the positive-parity phonons being responsible for the coupling between the selected particle states, the transitions between the bound and the continuum particle states originate from the coupling to the negative-parity phonons, such as the $3^{-}$ and $5^{-}$ phonons. Analogously to the bound  state $1\text{g}_{7/2}$, the continuum states $3\text{p}_{3/2}$ and $3\text{p}_{1/2}$  are either weakly or strongly fragmented in $^{68-72}\text{Ni}$ isotopes at $T>0$, and suddenly become good single-particle states in $^{74-78}\text{Ni}$ isotopes at $T=1$ MeV. For a quantitative comparison, Table \ref{Continuum states} shows the temperature evolution of the dominant fragments of the above mentioned continuum states in $^{70,74}\text{Ni}$ isotopes.  

It has been shown in the previous studies \cite{Litvinova2019h, Vaquero2020a} that the low-energy collective quadrupole $2^{+}$ and octupole $3^{-}$ phonons couple most strongly to the single-particle degrees of freedom. This is consistent with our qualitative study discussed in Section \ref{Toy models}, which emphasizes the importance of the low-energy phonons for the fragmentation of single-particle states. Recall that the phonon vertices $g^{m}_{k_{1}k_{2}}$ are the quantitative measure of the coupling strength for the given phonon mode $m$, and they are related to the phonon transition densities $\rho^{m}_{k_{1}k_{2}}$ by Eq. \eqref{Phonon vertices}. For each phonon mode $m$, at $T = 0$ the corresponding reduced transition probability $B_{m}(\omega)$ reads
\begin{equation}
\label{Reduced transition probability}B_{m}(\omega)=\left|\sum_{k_{1}k_{2}}V^{0}_{k_{1}k_{2}}\rho^{m}_{k_{1}k_{2}}(\omega)\right|^{2},
\end{equation}
where $V^{0}$ is the external field, which induces an excitation from the ground to the excited state $m$. At the pole of the corresponding response function $\omega=\omega_{m}$ the reduced transition probability $B_{m}(\omega_{m})$ is related to the strength function $S(\omega_{m})$ via \cite{LitvinovaRingTselyaev2007}
\begin{equation}
\label{Strength function at the pole}S(\omega_{m})=\lim_{\Delta\rightarrow+0}\frac{B_{m}(\omega_{m})}{\pi\cdot\Delta},
\end{equation}
where $\Delta$ is the smearing parameter. From Eqs. \eqref{Phonon vertices}, \eqref{Reduced transition probability}, and \eqref{Strength function at the pole}, one deduces that the larger the strength function $S(\omega_{m})$ at the pole, the larger are the matrix elements of the phonon vertices $g^{m}$ and the stronger is the PVC. A similar correlation holds for the case of finite temperature \cite{Wibowo2019a,Litvinova2021b}. Therefore, it is instructive to investigate the temperature dependence of the $2^{+}$ and $3^{-}$ phonon low-energy strength functions. Figs. \ref{Ni68_2+_Phonon}-\ref{Ni78_2+_Phonon} display this dependence for the quadrupole strength distributions in $^{68-78}\text{Ni}$. At $T=1$ MeV, one observes an attenuation of the low-energy $2^{+}$ phonon strength in $^{70}\text{Ni}$ before seeing it intensifying in $^{72-78}\text{Ni}$. In contrast, a gradual enhancement of the low-energy quadrupole phonon strengths has also been observed across the Ni isotopes at $T>1$ MeV.  These observations correlate with the behavior of the neutron hole $2\text{p}_{1/2}$, $1\text{f}_{5/2}$, and $2\text{p}_{3/2}$ states, which are  strongly fragmented in $^{74,76}\text{Ni}$ isotopes, whereas they are good single-particle states in $^{68-72}\text{Ni}$ isotopes, as follows from Figs. \ref{Ni68_RMF+PVC}-\ref{Ni76_RMF+PVC}. The temperature dependence of the octupole strength distributions across the Ni isotopes is again different, as shown in Figs. \ref{Ni68_3-_Phonon}-\ref{Ni78_3-_Phonon}. At $T=1$ MeV, one observes a slow, but steady attenuation of the low-lying phonon strengths throughout the Ni isotopes. At other temperatures, one observes rather a steady increase of the low-lying $3^{-}$ phonon strength. The attenuation of the low-lying phonon strengths at $T = 1$ MeV is responsible, for instance, for the change of the fragmentation pattern of the particle bound state $1\text{g}_{7/2}$ and of the continuum states $3\text{p}_{1/2}$ and $3\text{p}_{3/2}$, which transit from being fragmented states in $^{68-72}\text{Ni}$ to the good single-particle states in $^{74-78}\text{Ni}$.

\section{Summary and Outlook}

In this  work, we investigated the fragmentation patterns of the single-particle states in neutron-rich nuclei at finite temperature. The Dyson equation for the fermionic propagator with the energy-dependent mass operator (or dynamical self-energy) including the PVC mechanism was solved numerically in the basis of the thermal relativistic mean field for the even-even Ni isotopes with the atomic masses $A = 68-78$. As in the zero-temperature case, the dynamical self-energy at finite temperature is responsible for the fragmentation of the mean-field single-particle states, while finite temperature represents another dimension in the model parameter space to reveal the microscopic aspects of the particle-vibration coupling. Complete fragmented single-particle spectra in the 20 MeV window around the Fermi energies of the considered nuclei were extracted, and their temperature evolution was analyzed. 

To investigate the essential factors determining the fragmentation patterns of single-particle states, we examined the toy systems consisting of one, two and three single-particle states and one phonon. We found that the fragmentation is sensitive to such quantities as the phonon frequency, the PVC coupling strength, and the distance between the single-particle states. The sensitivity of the single-particle spectroscopic factors to these characteristics is quantified by varying them independently and with the temperature. These studies shed light on the fragmentation patterns obtained in the realistic calculations, in particular, we established how the temperature evolution of the phonon modes translates to the evolution  of the fragmentation patterns in Ni isotopes via the PVC mechanism.

The systematic studies presented in this work further advance the understanding of the behavior of atomic nuclei at extremal conditions. Here we investigated the extremes of the isospin and temperature, which are of prime importance for astrophysical modeling of cataclysmic events, such as the neutron star mergers and the supernova explosions. The nuclear single-particle properties in the astrophysical environments underly the behavior of the reaction rates, such as the neutron capture, beta decay and electron capture, which are pivotal for modeling the r-process nucleosynthesis and the core collapse of the supernovae. These rates require calculations of fermionic two-body propagators in correlated media in a similar manner, while the results obtained in this work can be directly used for studying the evolution of nuclear level densities with temperature, which are another important part of the nuclear physics input for the astrophysical modeling. Such studies will be addressed by future efforts. 

\section{Acknowledgment}
This work is partly supported by the US-NSF Career Grant PHY-1654379.
%


\bibliography{First_paper2021.bib,Bibliography_Aug2021.bib}

\end{document}